\documentclass[journal]{IEEEtran}
\usepackage{lscape}
\usepackage{array}
\usepackage{graphicx}
\usepackage{multicol}
\usepackage{multirow}
\usepackage{amssymb}
\usepackage{pifont}
\usepackage{tikz}
\usepackage{verbatim}
\usepackage{array}
\usepackage{longtable}
\usepackage{supertabular}
\usepackage{amsmath}
\usepackage{epstopdf}
\usepackage{verse}
\usepackage{cite}
\usepackage{color}
\usepackage{lettrine}
\usepackage{hyperref}
\usepackage{soul}
\usepackage{comment}
\usepackage{verbatim}
\usepackage{authblk}
\usepackage{caption}
\usepackage{subcaption}

\usepackage{booktabs}

\usepackage{tikz}
\usetikzlibrary{positioning,arrows.meta,calc}
\usepackage{fontawesome5} 

\usepackage{tikz}
\usetikzlibrary{matrix, positioning, arrows.meta, shapes.misc}

\tikzset{
  >=Latex, 
}

\usetikzlibrary{arrows.meta,calc,positioning,shapes.geometric}

\definecolor{navy}{HTML}{07546B}
\definecolor{cyan}{HTML}{008EAF}
\definecolor{cyanline}{HTML}{72CFE0}
\definecolor{cyanfill}{HTML}{F0FAFC}
\definecolor{red}{HTML}{B51F2E}
\definecolor{redicon}{HTML}{E83F4B}
\definecolor{redline}{HTML}{E7A6AC}
\definecolor{redfill}{HTML}{FFF7F7}
\definecolor{green}{HTML}{15883E}
\definecolor{graytext}{HTML}{4D6178}

\tikzset{
  card/.style={rounded corners=2.4mm, line width=.7pt, inner sep=0pt},
  bluecard/.style={card, draw=cyanline, fill=cyanfill},
  redcard/.style={card, draw=redline, fill=redfill},
  heading/.style={font=\bfseries\fontsize{12}{13}\selectfont, align=center},
  itemtitle/.style={font=\bfseries\fontsize{7.6}{8.7}\selectfont, align=center, text width=2.65cm},
  itemsub/.style={font=\fontsize{6.9}{8}\selectfont, align=center, text=graytext, text width=2.65cm},
  badge/.style={circle, minimum size=7.2mm, inner sep=0pt, text=white,
    font=\bfseries\fontsize{6.6}{7}\selectfont},
  evalhead/.style={font=\bfseries\fontsize{9.5}{10.5}\selectfont, text=navy, align=center},
  evaltext/.style={font=\fontsize{7.6}{9}\selectfont, text=navy, align=center, text width=4.8cm},
  flow/.style={-{Stealth[length=2.5mm,width=1.8mm]}, line width=1pt}
}

\newcommand{\blueitem}[5]{%
  \node[badge,fill=cyan] at (#1,#2) {#3};
  \node[itemtitle,text=navy,anchor=north] at (#1,#2-.52) {#4};
  \node[itemsub,anchor=north] at (#1,#2-1.43) {#5};
}
\newcommand{\reditem}[5]{%
  \node[badge,fill=redicon] at (#1,#2) {#3};
  \node[itemtitle,text=red,anchor=north] at (#1,#2-.52) {#4};
  \node[itemsub,anchor=north] at (#1,#2-1.43) {#5};
}

\usepackage{tikz}
\usetikzlibrary{matrix}

\usepackage{amsmath}
\usepackage{pifont}
\usepackage[normalem]{ulem} 
\usepackage{makecell}

\newcommand{\cmark}{\text{\ding{51}}}
\newcommand{\xmark}{\text{\ding{55}}}

\newcolumntype{a}{>{\columncolor{LightCyan}}c}

\begin{document}

\title{Trustworthy Distributed Machine Learning in Cloud-Edge-IoT Systems: A Heterogeneity-Aware Survey of Security, Privacy, and Robustness}

\author{Long Dang$^{*}$,
        Ali Alfatemi$^{*}$,
        Kaiqi~Xiong,~\IEEEmembership{Senior~Member,~IEEE,}
        Mohammed Aledhari,~\IEEEmembership{Senior~Member,~IEEE,}
        Abdellah~Chehri$^{*}$,~\IEEEmembership{Senior~Member,~IEEE,}
        Shawqi~Al-Maliki,
        and Mohamed~Rahouti,~\IEEEmembership{Member,~IEEE}

\thanks{\copyright~2026 IEEE. Personal use of this material is permitted. Permission from IEEE must be obtained for all other uses. This is the accepted manuscript of an article published in \emph{IEEE Internet of Things Journal}. DOI: \url{https://doi.org/10.1109/JIOT.2026.3742225}. The final authenticated version is available online at IEEE Xplore: \url{https://ieeexplore.ieee.org/document/11733665}. }

\thanks{L. Dang and K. Xiong are with Cyber Florida, University of South Florida, Tampa, FL, USA (e-mail: \{longdang; xiongk\}@usf.edu).}

\thanks{A. Alfatemi and M. Rahouti are with the Department of Computer and Information Science, Fordham University, New York, NY, USA (e-mail: \{aalfatemi; mrahouti\}@fordham.edu).}

\thanks{M. Aledhari is with the Data Science Department, University of North Texas, Denton, 76207, USA (e-mail: mohammed.aledhari@unt.edu).}

\thanks{A. Chehri is with the Department of Mathematics and Computer Science at the Royal Military College of Canada (RMC) in Kingston, Ontario, Canada (e-mail: chehri@rmc.ca).}

\thanks{S. Al-Maliki is with the Information and Computing Technology (ICT) Division, College of Science and Engineering, Hamad Bin Khalifa University, Doha 34110, Qatar (e-mail: shalmaliki@hbku.edu.qa).}

\thanks{Corresponding Author: \textbf{Abdellah Chehri, (e-mail: chehri@rmc.ca)}}

\thanks{$^{*}$These authors contributed equally to this work.}
}

\maketitle

\begin{abstract}
Distributed machine learning (DML), including conventional distributed training and federated learning (FL), enables collaborative model development across cloud-edge-IoT systems without fully centralizing raw data. However, distributing data, computation, updates, and orchestration increases the attack surface and introduces coupled challenges across privacy, robustness, reliability, fairness, and governance. This survey presents a security-centered, heterogeneity-aware roadmap for trustworthy DML, with FL treated as a prominent high-risk case. We first characterize the assets, adversaries, trust assumptions, and coordination patterns that define DML/FL security boundaries. We then analyze how statistical, system, and trust heterogeneity, including non-IID data, resource asymmetry, intermittent participation, and cross-silo/cross-device trust differences, affect convergence, fairness, attack detectability, and defense reliability. Next, we synthesize attack-defense interactions across poisoning, backdoors, Byzantine manipulation, privacy leakage, model extraction, network/orchestration attacks, Sybil behavior, and strategic participation. We further compare layered defenses, including robust aggregation, drift-aware optimization, differential privacy, secure aggregation, homomorphic encryption, secure multiparty computation, trusted execution environments, watermarking, provenance, authenticated communication, and monitoring. Finally, we discuss deployment, evaluation, governance, and open research guidelines, emphasizing reproducible, non-independent and identically distributed (non-IID) partitions, attack-aware benchmarks, privacy budgets, attack success rate, client-level disparity, latency, communication, memory, energy/carbon overhead, and auditable lifecycle artifacts.
\end{abstract}
\IEEEpeerreviewmaketitle

\section{Introduction} \label{introduction}

\IEEEPARstart{D}{istributed} machine learning (DML) has become a core paradigm for training models across cloud, edge, and Internet of Things (IoT) environments. In these settings, data are often generated by heterogeneous devices, organizations, sensors, and embedded platforms, making centralized collection costly, privacy-sensitive, or infeasible. DML, including conventional distributed training and federated learning (FL), enables collaborative model development while keeping computation and data distributed across resource-diverse participants \cite{verbraeken2020survey, mcmahan2017communication, li2020federated}.

However, distributing learning also expands the attack surface. Model updates, gradients, aggregation logic, communication channels, client identities, and orchestration infrastructure may all become targets. These risks are amplified in IoT-oriented deployments, where devices often operate with limited compute, energy, and bandwidth resources, and with limited security hardening. In addition, statistical heterogeneity, especially non-independent and identically distributed (non-IID) data, can induce client drift, slow convergence, widen client-level performance disparities, and weaken assumptions behind many robust aggregation and privacy-preserving mechanisms \cite{li2020federated, kairouz2021advances}. As a result, poisoning, backdoor insertion, Byzantine update manipulation, gradient leakage, model extraction, and network-level attacks must be studied jointly with system and data heterogeneity \cite{Yuan2019AdversarialLearning}.

This survey takes a security-centered view of DML in cloud-edge-IoT systems. Rather than providing tutorial-level coverage of generic distributed computing platforms, basic optimization, or cloud ML services, we focus on how threat models, heterogeneity, privacy mechanisms, robustness limits, and deployment constraints interact. FL is treated as a prominent and high-risk instance of DML because it combines distributed data ownership, partial trust, client churn, and non-IID data. The goal is to provide a heterogeneity-aware roadmap for the design, evaluation, and deployment of trustworthy DML systems.

\begin{table*}[h]
\scriptsize
\centering
\setlength{\tabcolsep}{2.8pt}
\renewcommand{\arraystretch}{1.12}
\caption{Comparison of representative surveys related to secure distributed and federated machine learning. 
\cmark\ indicates substantial coverage, $\triangle$ indicates partial or secondary coverage, and \xmark\ indicates limited or no meaningful coverage. 
The comparison reflects relative emphasis based on survey scope rather than an exhaustive claim about all topics mentioned in each paper}
\label{tab:related_surveys}
\begin{tabular}{|p{1.55cm}|p{0.5cm}|p{0.8cm}|p{1.35cm}|p{1.4cm}|p{1.15cm}|p{1.25cm}|p{1.35cm}|p{6.8cm}|}
\hline
\textbf{Survey} &
\textbf{Year} &
\textbf{DML / FL} &
\textbf{Threat / Trust Model} &
\textbf{Heterogeneity} &
\textbf{Attack-Defense} &
\textbf{Privacy / Protected Comp.} &
\textbf{Deploy. / Eval. / Gov.} &
\textbf{Main Emphasis} \\
\hline \hline

Verbraaken et al.~\cite{verbraeken2020survey} &
2020 & \cmark & \xmark & $\triangle$ & \xmark & \xmark & $\triangle$ &
Broad distributed deep learning systems survey focused mainly on scalability, orchestration, and infrastructure rather than security. \\
\hline

Li et al.~\cite{li2020federated} &
2020 & \cmark & $\triangle$ & \cmark & $\triangle$ & $\triangle$ & $\triangle$ &
Foundational FL survey covering architectures, statistical heterogeneity, communication efficiency, optimization, and privacy-related challenges. \\
\hline

Yang et al.~\cite{yang2020adversary} &
2020 & \cmark & $\triangle$ & \cmark & $\triangle$ & $\triangle$ & $\triangle$ &
Early FL survey addressing concepts, communication settings, application domains, and adversarial considerations. \\
\hline

Kairouz et al.~\cite{kairouz2021advances} &
2021 & \cmark & $\triangle$ & \cmark & $\triangle$ & $\triangle$ & $\triangle$ &
Landmark FL survey spanning optimization, personalization, privacy, fairness, robustness, and open problems, but not centered on a security-systems roadmap. \\
\hline

Rodr\'iguez-Barroso et al.~\cite{rodriguez2023survey} &
2023 & \cmark & \cmark & $\triangle$ & \cmark & $\triangle$ & $\triangle$ &
Comprehensive FL threats survey with attack/defense taxonomies and empirical analysis across privacy and integrity threats. \\
\hline

Zhang et al.~\cite{zhang2024survey} &
2024 & \cmark & $\triangle$ & $\triangle$ & \cmark & $\triangle$ & $\triangle$ &
Trustworthy FL survey covering privacy, security, fairness, explainability, and accountability under a broad trustworthiness lens. \\
\hline

Chai et al.~\cite{chai2024survey} &
2024 & \cmark & \xmark & \cmark & $\triangle$ & \xmark & \cmark &
Survey focused on FL evaluation, benchmarking, metrics, fairness, and performance assessment rather than security mechanisms. \\
\hline

Wan et al.~\cite{wan2024data} &
2024 & \cmark & $\triangle$ & $\triangle$ & \cmark & \xmark & $\triangle$ &
Security-focused survey of backdoor attacks and defenses for FL in wireless communication settings. \\
\hline

Nguyen et al.~\cite{nguyen2024backdoor} &
2024 & \cmark & $\triangle$ & \xmark & \cmark & \xmark & $\triangle$ &
Systematic review of backdoor attacks and defenses in FL, emphasizing attack taxonomy and mitigation trends. \\
\hline

Zhao et al.~\cite{10.1145/3724113} &
2025 & \cmark & $\triangle$ & $\triangle$ & $\triangle$ & \cmark & $\triangle$ &
Dedicated survey on FL privacy attacks and defenses, including inference risks, mitigation methods, applications, and policy considerations. \\
\hline

Feng et al.~\cite{feng2025survey} &
2025 & \cmark & \cmark & \xmark & \cmark & $\triangle$ & \xmark &
Security-focused FL survey addressing backdoor, Byzantine, and adversarial threats with corresponding defense taxonomies. \\
\hline

Li et al.~\cite{li2025backdoor} &
2025 & \cmark & $\triangle$ & \xmark & \cmark & \xmark & \xmark &
Focused survey of backdoor attacks and defenses in FL, including attack settings, triggers, and evaluation criteria. \\
\hline

Zhang et al.~\cite{zhang2025review} &
2025 & \cmark & $\triangle$ & \xmark & $\triangle$ & \cmark & $\triangle$ &
Survey dedicated to secure aggregation in FL, comparing protocol families, efficiency, and deployment trade-offs. \\
\hline

Uddin et al.~\cite{uddin2025systematic} &
2025 & \cmark & $\triangle$ & $\triangle$ & \cmark & \xmark & $\triangle$ &
Systematic/thematic review of robust FL methods, including client- and server-side robustness strategies, applications, and gaps. \\
\hline

Mahmud et al.~\cite{mahmud2026survey} &
2026 & \cmark & $\triangle$ & $\triangle$ & \cmark & $\triangle$ & $\triangle$ &
Broad FL survey covering architectures, synchronization, and security threats, but without a heterogeneity-centered security synthesis. \\
\hline

\textbf{This survey} &
\textbf{2026} & \cmark & \cmark & \cmark & \cmark & \cmark & \cmark &
Security-centered and heterogeneity-aware survey of DML that connects threat models, trust boundaries, heterogeneity, attack-defense interactions, protected computation, evaluation practice, and deployment/governance considerations. \\
\hline

\end{tabular}
\end{table*}

\subsection{Related Surveys}

Prior surveys have examined DML and FL from several complementary perspectives, including distributed training architectures, FL optimization, statistical heterogeneity, privacy-preserving learning, adversarial robustness, trustworthiness, and evaluation practice \cite{verbraeken2020survey, li2020federated, yang2020adversary, kairouz2021advances, chai2024survey}. Other surveys focus more directly on FL security, including privacy and integrity attacks, backdoor and Byzantine threats, secure aggregation, and broader trustworthy-FL concerns \cite{rodriguez2023survey, zhang2024survey, nguyen2024backdoor, 10.1145/3724113, feng2025survey, li2025backdoor, zhang2025review, mahmud2026survey}. Robust FL and wireless/backdoor-oriented surveys further provide focused taxonomies for specific subareas \cite{uddin2025systematic, wan2024data}.

As summarized in Table~\ref{tab:related_surveys}, existing work provides valuable coverage of individual aspects of DML and FL. However, fewer surveys synthesize the interaction among statistical and system heterogeneity, Byzantine and backdoor robustness, privacy mechanisms such as differential privacy and secure aggregation, stronger protected-computation techniques such as homomorphic encryption (HE), secure multiparty computation (MPC), and trusted execution environments (TEEs), and deployment-oriented concerns such as reproducibility, governance, and operational security.

This survey addresses that gap by organizing secure DML around four connected questions: what assets and trust boundaries must be protected; how IoT-oriented heterogeneity changes learning and security assumptions; how attacks and defenses interact under realistic deployment constraints; and how systems should be evaluated beyond predictive accuracy.

\begin{figure}[h]
    \centering
    \resizebox{!}{0.5\textwidth}{
\begin{tikzpicture}[x=1cm,y=1cm]

\path[bluecard] (0,7.0) rectangle (5.55,18.2);
\path[redcard]  (5.90,7.0) rectangle (11.45,18.2);
\path[bluecard] (11.80,7.0) rectangle (17.35,18.2);

\node[heading,text=navy] at (2.775,17.45) {Non-IID\\Heterogeneity};
\node[heading,text=red]  at (8.675,17.45) {Attack Surface};
\node[heading,text=navy] at (14.575,17.45) {Defense Portfolio};

\blueitem{1.45}{15.75}{LS}{Label skew}{$P_k(y)$}
\blueitem{4.10}{15.75}{FS}{Feature/covariate\\shift}{$P_k(x)$}
\blueitem{1.45}{12.95}{CS}{Concept shift}{$P_k(y\mid x)$}
\blueitem{4.10}{12.95}{QS}{Quantity skew}{$|D_k|$}
\blueitem{2.775}{9.85}{ST}{Spatiotemporal skew}{}

\reditem{7.25}{15.75}{BP}{Byzantine/poisoning}{including backdoors}
\reditem{10.10}{15.75}{GI}{Gradient inversion}{DLG / iDLG / IG}
\reditem{7.25}{12.75}{MI}{Membership inference}{}
\reditem{10.10}{12.75}{ME}{Model extraction/stealing}{}
\reditem{7.25}{9.65}{NT}{Network tampering / MITM}{}
\reditem{10.10}{9.65}{SY}{Sybil \& free-rider behavior}{}

\blueitem{13.15}{15.75}{HO}{Heterogeneity-aware optimization}{FedProx, SCAFFOLD}
\blueitem{16.00}{15.75}{RA}{Robust aggregation}{Median, Krum, RFA}
\blueitem{13.15}{12.65}{SA}{Secure aggregation}{SA}
\blueitem{16.00}{12.65}{DP}{Differential privacy}{DP-SGD, PATE}
\blueitem{13.15}{9.80}{CM}{Cryptographic ML}{HE / MPC / TEE}
\blueitem{16.00}{9.80}{IP}{IP protection}{Watermarking}
\node[badge,fill=cyan] at (12.75,7.55) {CV};
\node[font=\bfseries\fontsize{7.6}{8.7}\selectfont,text=navy,anchor=west]
  at (13.32,7.6) {Client vetting};
\node[font=\fontsize{6.8}{7.8}\selectfont,text=graytext,anchor=west]
  at (13.32,7.27) {Anomaly scoring; attestation};

\node[font=\bfseries\fontsize{7.8}{9}\selectfont,text=navy] at (2.775,6.05) {report heterogeneity};
\draw[flow,cyan] (2.775,5.72) -- (2.775,5.16);

\node[font=\bfseries\fontsize{8.2}{9.5}\selectfont,text=redicon] at (7.55,5.62) {$\rightarrow$ exacerbates};
\node[font=\bfseries\fontsize{8.2}{9.5}\selectfont,text=green] at (10.30,5.62) {$\rightarrow$ mitigate via};

\node[font=\bfseries\fontsize{7.8}{9}\selectfont,text=navy] at (14.575,6.05) {report \& certify};
\draw[flow,cyan] (14.575,5.72) -- (14.575,5.16);

\path[bluecard] (0,0) rectangle (17.35,4.70);
\node[heading,text=navy] at (8.675,4.03) {Evaluation \& Standards};

\node[evalhead] at (2.90,2.65) {Metrics};
\node[evaltext,anchor=north] at (2.90,2.20)
  {Accuracy, ASR, per-client fairness\\DP $(\epsilon,\delta)$, communication, energy};

\draw[cyanline,line width=.6pt] (5.75,1.00) -- (5.75,3.15);
\node[evalhead] at (8.675,2.65) {Benchmarks};
\node[evaltext,anchor=north] at (8.675,2.20)
  {Standardized non-IID partitions and seeds\\Public, reproducible attack suites};

\draw[cyanline,line width=.6pt] (11.60,1.00) -- (11.60,3.15);
\node[evalhead] at (14.45,2.65) {Governance};
\node[evaltext,anchor=north] at (14.45,2.20)
  {Model/data cards and threat models\\Privacy accounting; NIST AI RMF; EU AI Act};

\end{tikzpicture}
}

\caption{Roadmap for heterogeneity-aware trustworthy DML in cloud-edge-IoT systems. Statistical non-IID heterogeneity can exacerbate poisoning, Byzantine manipulation, privacy leakage, model extraction, network tampering, and Sybil/free-rider behavior. These risks motivate a layered defense portfolio spanning heterogeneity-aware optimization, robust and secure aggregation, differential privacy, protected computation, intellectual property (IP) protection, and client vetting. Evaluation should report utility, attack success rate, client-level fairness, privacy budgets, communication/energy overhead, standardized non-IID partitions, public attack suites, and governance artifacts.}
\label{fig:taxonomy}
\end{figure}
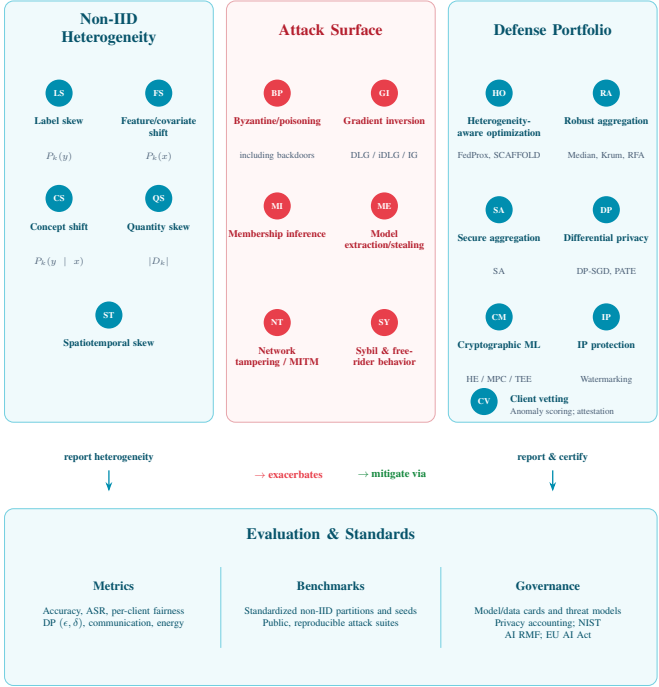

\subsection{Contributions}

This survey examines trustworthy DML under realistic IoT-oriented conditions involving non-IID data, resource asymmetry, partial trust, and adaptive adversaries. Its main contributions are:

\begin{itemize}
    \item Threat-centered DML taxonomy: We systematize the attack surface across distributed and FL, including poisoning, backdoors, Byzantine manipulation, privacy leakage, model extraction, and system/network-level attacks.

    \item Heterogeneity-aware security analysis: We analyze how statistical, system, and trust heterogeneity affect convergence, fairness, attack detectability, and the reliability of common defenses.

    \item Defense trade-off synthesis: We compare robust aggregation, differential privacy, secure aggregation, HE, MPC, TEEs, watermarking, fingerprinting, and provenance mechanisms in terms of accuracy, privacy, latency, communication, trust assumptions, and deployment complexity.

    \item Deployment and evaluation guidance: We consolidate IoT-oriented evaluation practices, including attack-aware benchmarking, reproducible non-IID partitions, privacy budgets, attack success rate, client-level disparity, communication overhead, latency, energy, and governance-oriented reporting.
\end{itemize}

Fig.~\ref{fig:taxonomy} summarizes the organizing logic of this survey. It illustrates how non-IID data and statistical heterogeneity should be explicitly reported, as they can exacerbate attacks on integrity, privacy, network, and strategic participation. The figure further maps this expanded attack surface to a layered defense portfolio and to evaluation and governance practices needed for trustworthy DML in cloud-edge-IoT systems.

\subsection{Paper Organization}

The remainder of this paper is organized as follows. Section~\ref{sec:threat_model} defines the DML/FL threat model and trust boundaries. Section~\ref{sec:heterogeneity} analyzes statistical, system, and trust heterogeneity in IoT-oriented DML. Section~\ref{sec:attack_surface} examines major attack surfaces and attack-defense interactions. Section~\ref{sec:defenses} reviews layered defense mechanisms and their trade-offs. Section~\ref{sec:deployment_guidelines} presents deployment, evaluation, and governance guidelines. Section~\ref{sec:open} discusses open research directions, and Section~\ref{sec:conclusion} concludes the paper.

\section{Threat Model and DML Trust Boundaries}
\label{sec:threat_model}

Distributed machine learning (DML) distributes data, computation, model updates, and orchestration logic across workers, edge devices, organizational silos, and cloud services. Conventional distributed training often operates within a single administrative domain. In contrast, FL and IoT-oriented DML commonly involve partially trusted clients, intermittent participation, heterogeneous resources, and non-IID local data \cite{verbraeken2020survey, li2020federated, kairouz2021advances}. From a security perspective, the central issue is therefore not only how models are optimized, but where trust boundaries arise and what information crosses them.

\subsection{Assets, Adversaries, and Trust Assumptions}

The primary assets in DML include raw local data, labels, gradients, model updates, global model parameters, aggregation rules, client identities, communication metadata, audit logs, and deployed model artifacts. In regulated IoT domains such as healthcare, finance, industrial automation, and smart cities, these assets may encode sensitive personal, operational, or proprietary information. Protecting them requires simultaneous attention to confidentiality, integrity, availability, robustness, and model ownership.

Adversaries may be external or internal. External adversaries can eavesdrop on, replay, or tamper with model traffic; query deployed models to extract or infer membership; or inject malicious inputs at inference time. Internal adversaries include compromised clients, malicious data owners, curious aggregators, compromised gateways, free-riding participants, and Sybil clients that create multiple identities to bias training or incentives \cite{fung2018mitigating, fang2020local, fraboni2021free}. Aggregators and cloud providers may be trusted, honest-but-curious, or partially compromised, depending on the deployment model. These assumptions determine whether defenses such as secure aggregation, differential privacy (DP), HE, secure MPC, TEEs, or robust aggregation are appropriate.

\subsection{DML/FL Coordination Patterns as Security Boundaries}

Coordination architecture determines which entities observe local updates, where aggregation occurs, and how failures or attacks propagate. In parameter-server training, workers send gradients or updates to one or more central servers, concentrating trust and making the servers potential bottlenecks, privacy-leakage points, and targets for poisoning \cite{186214, li2020federated}. Collective communication mechanisms such as AllReduce reduce dependence on a persistent parameter server in trusted clusters. However, they still require authenticated group membership, secure transport, and resilience to faulty participants \cite{mayer2020scalable, NVIDIADeveloper}.

Data-parallel, model-parallel, and hybrid-parallel execution also expose different security surfaces. Data parallelism communicates gradients or model deltas from local data shards, making it especially relevant to poisoning, gradient leakage, and robust aggregation. Model and pipeline parallelism communicate activations, gradients, and intermediate states across model partitions, thereby creating additional dependency paths and increasing the risk of fault propagation \cite{shoeybi2019megatron, huang2019gpipe, narayanan2021efficient}. Synchronization further shapes security: synchronous training simplifies aggregation analysis but is vulnerable to stragglers and denial-of-service behavior, whereas asynchronous or bounded-staleness training improves throughput under heterogeneity but introduces stale updates, replay risk, and weaker anomaly attribution \cite{chen2016revisiting, sapio2021scaling, cui2014exploiting}.

\subsection{IoT-Specific Exposure Points: Edge Clients, Gateways, Aggregators, and Cloud Services}

IoT-oriented DML adds exposure points not typically present in controlled data-center training. Edge clients may have weak hardware isolation, limited patching, intermittent power, and constrained storage or compute. Gateways and roadside/industrial edge nodes may aggregate updates from multiple devices and therefore become attractive targets for poisoning, traffic analysis, or model substitution \cite{mahadevappa2024analyzing}. Cloud aggregators and ML-as-a-service platforms provide scalability but also introduce concerns related to multi-tenancy, data residency, access control, and model extraction \cite{liu2025data}.

Beyond conventional FL threat models, IoT deployments introduce physical and operational constraints that directly shape security design \cite{quan2025federated}. Clients may be battery-powered, intermittently connected, memory-limited, weakly patched, or physically accessible to an adversary. Sensing modalities may also differ across devices, e.g., cameras, accelerometers, medical wearables, industrial vibration sensors, vehicular LiDAR/radar, or environmental sensors, producing modality-specific drift and uneven update quality. In addition to network-level tampering, IoT systems are exposed to physical-layer disruption, sensor spoofing, jamming, device theft, firmware compromise, gateway compromise, and stale or replayed telemetry. These risks make it insufficient to assume that every participating client is continuously available, computationally capable, or semantically trustworthy. Secure IoT DML therefore requires explicit placement decisions across the client, gateway, edge, and cloud layers: lightweight filtering and attestation near devices; buffering, rate control, and anomaly screening at gateways; aggregation, privacy accounting, and provenance at edge/cloud servers; and rollback or fallback policies for safety-critical deployments \cite{goudarzi2021distributed}.

Accordingly, IoT DML should be analyzed as a layered system comprising client-side data and local training; gateway-level filtering and buffering; server-side aggregation and model governance; and cloud/edge deployment of inference services. Each layer can alter the feasibility and cost of privacy, robustness, and cryptographic defenses \cite{al2024computing}.

\section{Heterogeneity Dimensions in IoT-Oriented DML}
\label{sec:heterogeneity}

Heterogeneity is a structural property of IoT-oriented DML rather than an implementation detail. It affects convergence, fairness, privacy leakage, attack detectability, and the reliability of defense mechanisms. We distinguish three interacting forms: statistical heterogeneity, system heterogeneity, and trust heterogeneity.

\subsection{Statistical Heterogeneity: Non-IID Data, Client Drift, and Fairness Disparity}

In distributed and FL, client $k$ typically holds data from a local distribution $\mathcal{D}_k$ and optimizes
\[
f_k(w) := \mathbb{E}_{(x,y)\sim \mathcal{D}_k}[\ell(w;x,y)] ,
\]
while global training seeks
\[
f(w) := \sum_{k=1}^{K} p_k f_k(w), \qquad 
p_k \ge 0,\quad \sum_{k=1}^{K} p_k = 1 .
\]
Non-IID data means that the local distributions $\mathcal{D}_1,\ldots,\mathcal{D}_K$ differ in feature marginals $P(x)$, label marginals $P(y)$, conditional distributions $P(y|x)$, or data volume. Common forms include label skew, feature or covariate shift, concept shift, quantity skew, and spatiotemporal skew \cite{li2021fedbn, li2021model, jimenez2025thorough, kairouz2021advances}.

\begin{figure}[h]
  \centering
  \begin{subfigure}[b]{0.24\textwidth}
    \centering
    \includegraphics[width=\linewidth]{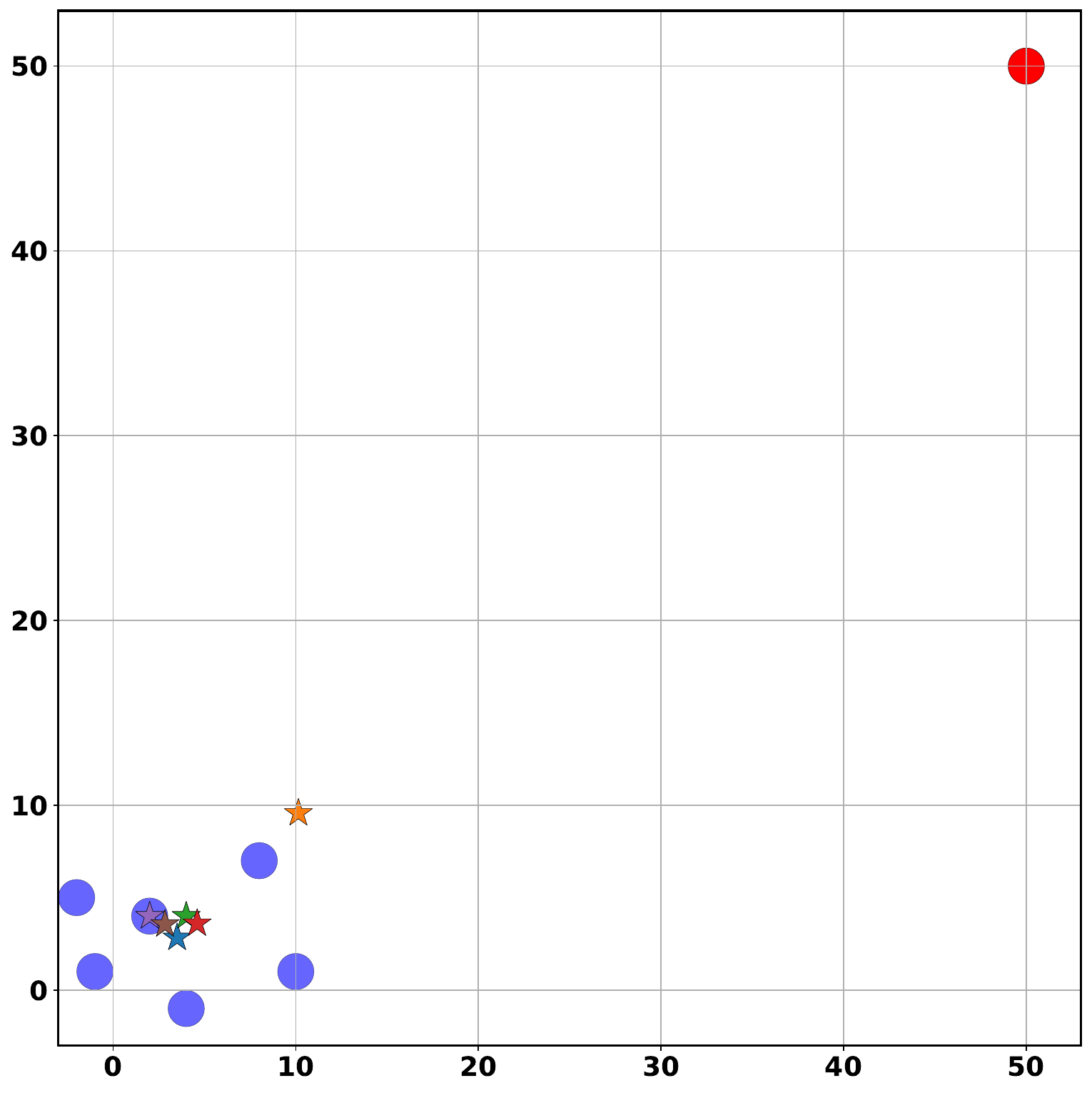}
    \caption{IID data.}
    \label{fig:robust_methods_IID}
  \end{subfigure}\hfill
  \begin{subfigure}[b]{0.24\textwidth}
    \centering
    \includegraphics[width=\linewidth]{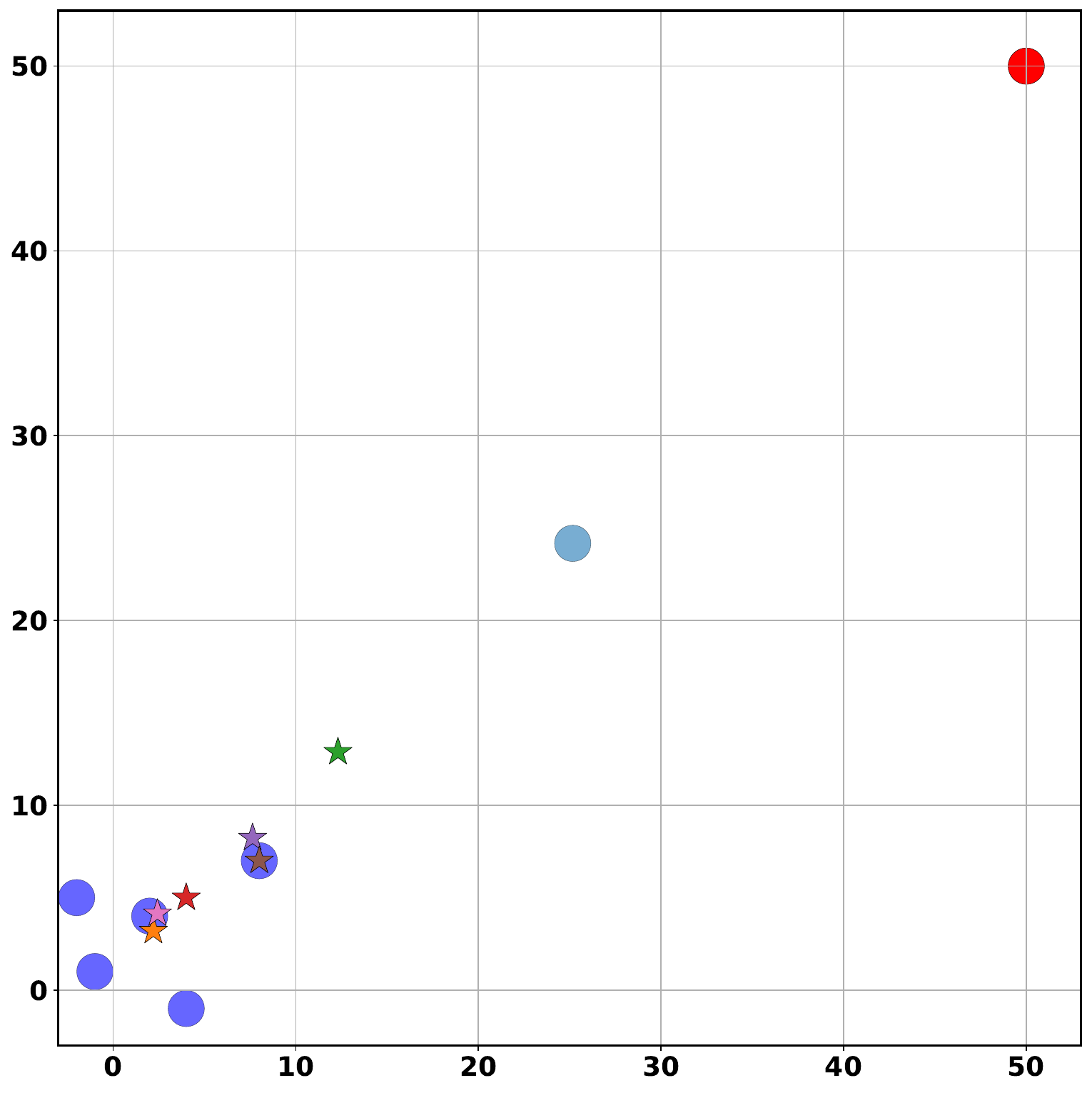}
    \caption{Non-IID degradation.}
    \label{fig:robust_methods_nonIID}
  \end{subfigure}\hfill
  \begin{subfigure}[b]{0.24\textwidth}
    \centering
    \includegraphics[width=\linewidth]{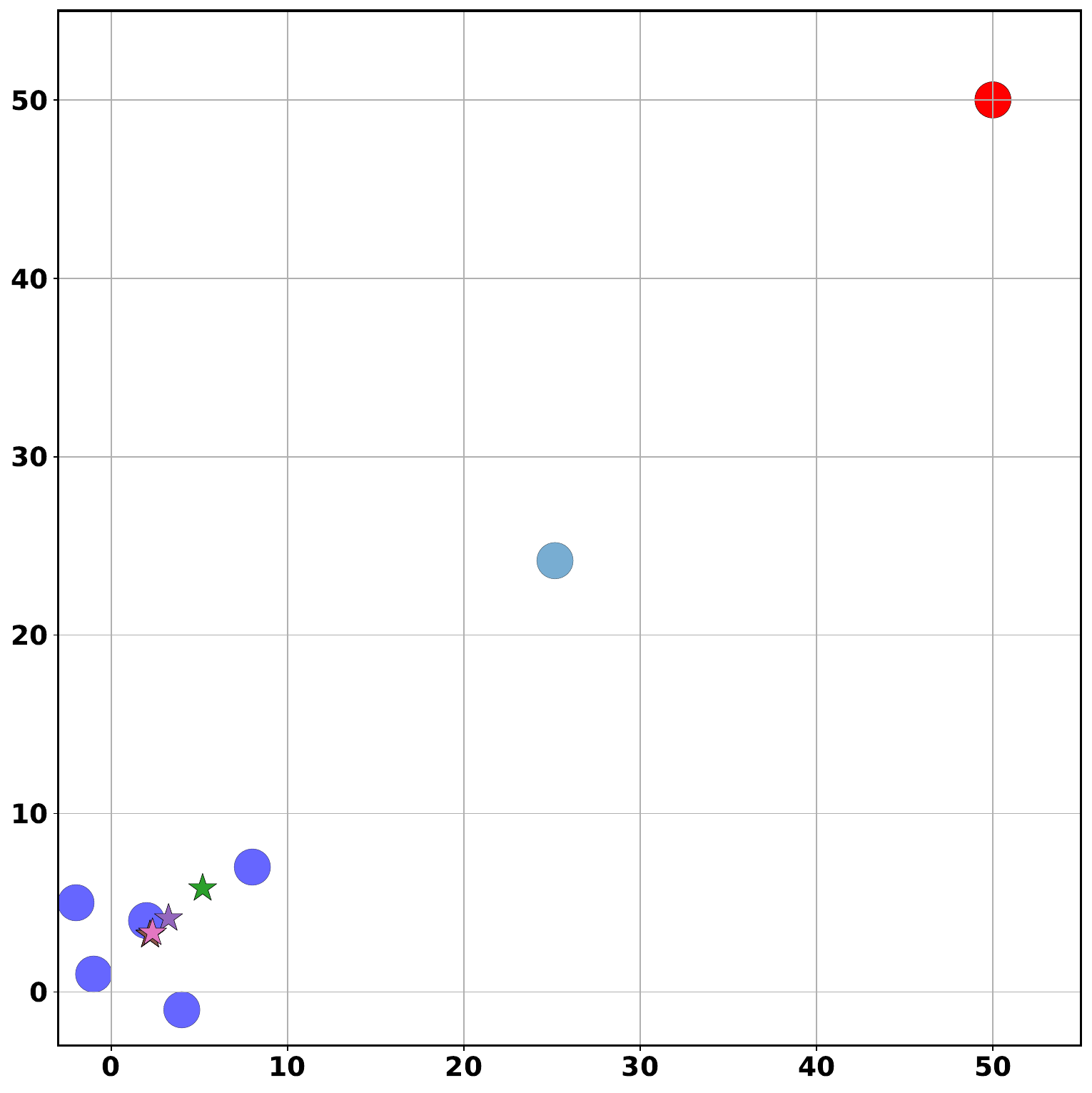}
    \caption{Nearest-neighbor mixing~\cite{allouah2023fixing}.}
    \label{fig:preprocessing1_nonIID}
  \end{subfigure}\hfill
  \begin{subfigure}[b]{0.24\textwidth}
    \centering
    \includegraphics[width=\linewidth]{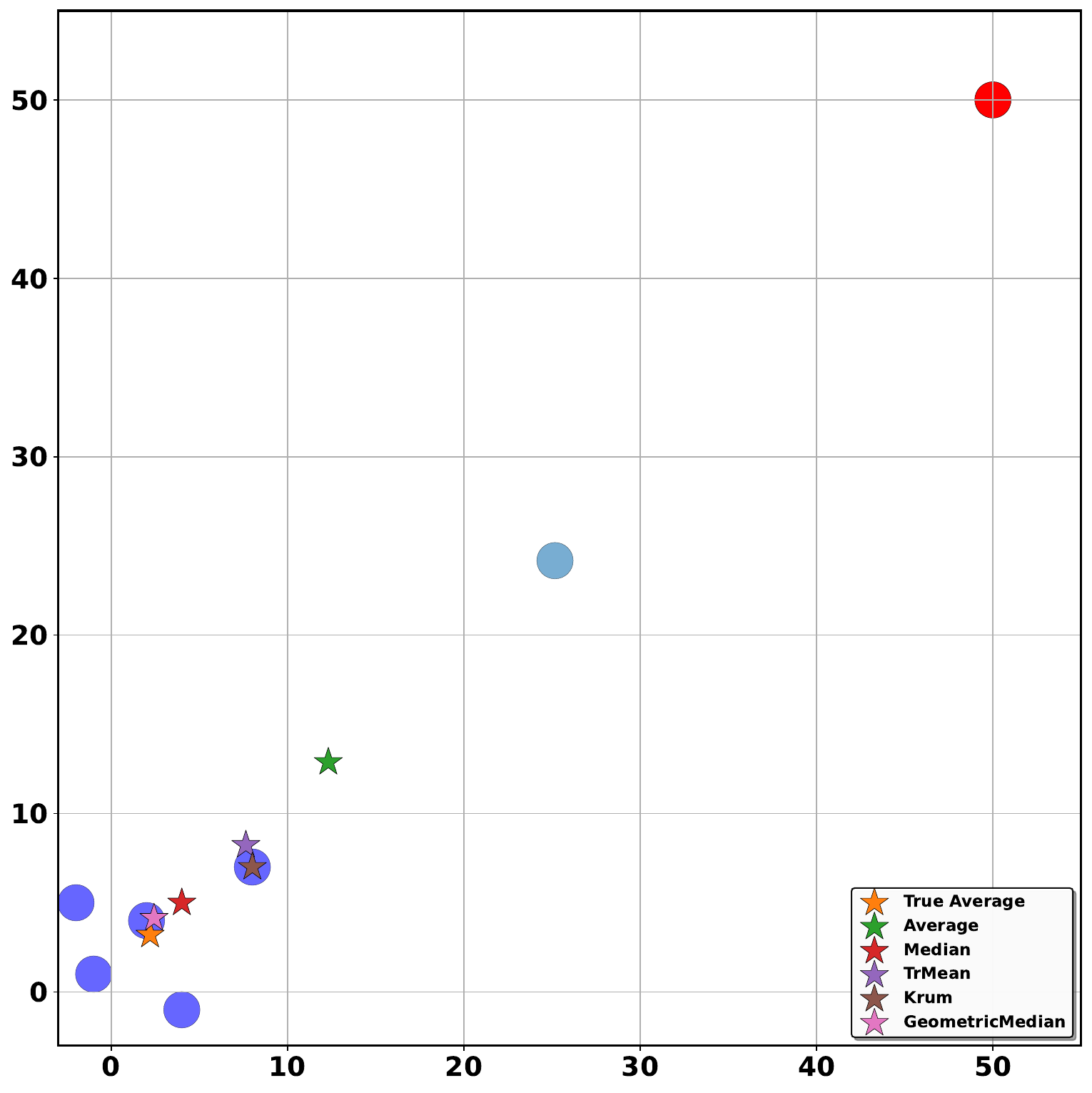}
    \caption{Bucketing~\cite{karimireddy2020byzantine}.}
    \label{fig:preprocessing2_nonIID}
  \end{subfigure}
  \caption{Effect of heterogeneity on Byzantine-robust aggregation. Robust rules such as median, trimmed mean, Krum, and geometric median are more reliable when benign updates are concentrated; under non-IID data, honest dispersion can reduce attack detectability. Preprocessing methods such as nearest-neighbor mixing and bucketing can partially mitigate this effect. Data points are from~\cite{yang2020adversary}, implemented using~\cite{gonzalez2025byzflresearchframeworkrobust}.}
  \label{fig:byzantine}
\end{figure}

Non-IID data induces gradient dissimilarity between local and global objectives. When clients perform multiple local updates before synchronization, local models may drift toward client-specific optima, slowing or destabilizing convergence and increasing client-level performance disparities \cite{karimireddy2020scaffold, li2020federated, zhao2018federated}. See Fig.~\ref{fig:robust_methods_nonIID},~\ref{fig:preprocessing1_nonIID}, and~\ref{fig:preprocessing2_nonIID} for the impact of non-IID data on robust aggregation methods and potential solutions. This is not only an optimization concern. Many Byzantine-robust aggregation rules assume that benign updates are relatively concentrated; under severe heterogeneity, honest updates may already appear as outliers, making malicious updates harder to detect \cite{pillutla2022robust, li2023experimental}.

Non-IID evaluations should therefore report both the partition protocol and explicit measures of heterogeneity. A common approach is Dirichlet label partitioning, in which each client's class proportions are sampled from a Dirichlet distribution with concentration parameter $\alpha$; smaller $\alpha$ values yield more highly skewed client distributions, whereas larger $\alpha$ values yield more nearly uniform distributions. Common heterogeneity metrics include Earth Mover's Distance (EMD), the minimum cost of transporting probability mass between distributions; Hellinger distance, a bounded measure of separation between probability distributions; Jensen-Shannon (JS) divergence, a symmetric and bounded variant of Kullback-Leibler divergence; and Kullback-Leibler (KL) divergence, a directed measure of relative entropy. Naturally heterogeneous benchmarks include \textsc{LEAF}, FEMNIST, and Shakespeare \cite{caldas2018leaf, zhao2018federated, jimenez2025thorough, chen2025effective}. These partitioning approaches and metrics are discussed further in the next sections.

\subsection{System Heterogeneity: Device, Bandwidth, Energy, and Participation Asymmetry}

IoT DML systems also exhibit strong system heterogeneity. Clients may differ in processor type, memory, battery level, sensor modality, local storage, communication bandwidth, and reliability. These differences affect which clients participate, how often they synchronize, and how expensive each defense becomes. For example, secure aggregation, MPC, and HE can increase communication or computation overhead; DP can reduce utility under small local datasets; and robust aggregation can require additional client scoring, clipping, or auditing.

System heterogeneity also affects availability and reliability. Stragglers, dropped clients, gateway failures, software incompatibilities, and unstable networks may delay or bias aggregation. Classical mechanisms such as checkpointing, rollback, replication, speculative execution, and elastic scheduling remain relevant, but they must be adapted to continuously evolving model states and repeated communication rounds \cite{mayer2020scalable, myllyaho2022misbehaviour}. Algorithm-based fault tolerance and Byzantine-resilient aggregation further link reliability and security, as corrupted updates can arise from either accidental faults or malicious behavior \cite{roffe2020evaluation, pillutla2022robust, li2023experimental}.

Energy and latency are also first-class constraints. A statistically effective defense may be infeasible on battery-powered sensors, mobile devices, or real-time industrial/vehicular platforms. Thus, secure IoT DML requires reporting on communication, latency, memory, and energy overhead, as well as accuracy and robustness.

\subsection{Trust Heterogeneity: Cross-Device, Cross-Silo, and Multi-Tenant Settings}

Trust assumptions vary substantially across deployment settings. In cross-silo FL, participants such as hospitals, banks, or factories may be identifiable institutions with contractual obligations but strict data-governance requirements \cite{kyriakidou2026toward}. In cross-device FL, participation is large-scale, intermittent, and difficult to vet \cite{pfeiffer2023federated} fully. In multi-tenant cloud ML, the provider, tenants, APIs, and infrastructure introduce additional risks related to access control, isolation, and model extraction.

These trust differences directly affect defense selection. Cross-silo deployments may justify TEEs, MPC, contractual audit logs, and stronger identity management \cite{korneev2025survey}. Cross-device deployments often require secure aggregation, capped client sampling, Sybil resistance, DP, and lightweight anomaly scoring. Multi-tenant MLaaS settings require authenticated interfaces, rate limiting, output control, watermarking, and provenance mechanisms. A defense that is suitable under one trust model may be ineffective or too costly under another.

Taken together, these three dimensions of heterogeneity provide the analytical lens for the remainder of the survey. Statistical heterogeneity affects convergence, benign-update dispersion, fairness, and poisoning detectability; system heterogeneity determines whether defenses are computationally and energetically feasible; and trust heterogeneity determines which privacy, aggregation, cryptographic, provenance, and governance mechanisms are appropriate. Sections~IV-VI therefore analyze attacks, defenses, and evaluation practices through these three coupled dimensions rather than treating heterogeneity as a standalone optimization issue.

\section{Attack Surface and Attack-Defense Interactions}
\label{sec:attack_surface}

DML attacks target the training pipeline, inference interface, communication substrate, identity layer, and deployed model artifact. Their impact extends beyond aggregate accuracy to safety, privacy, fairness, intellectual property, operational continuity, and regulatory exposure. To move beyond a descriptive taxonomy, Table~\ref{tab:attack_defense_analysis} compares major attack classes in terms of threat assumptions, adversary capabilities, attack objectives, affected pipeline stages, candidate defenses, and residual defense limitations.

The table should be read as an analytical roadmap for the subsections that follow. Section~\ref{sec:poisoning_byzantine} first examines training-time integrity attacks, including poisoning, backdoors, and Byzantine manipulation. The subsequent subsections address privacy and IP leakage; network/orchestration and strategic-participation attacks; and, finally, the cross-cutting role of heterogeneity in shaping attack detectability and defense reliability.

\begin{table*}[h]
\centering
\scriptsize
\caption{Analytical comparison of major DML/FL attack classes and defense-composition limits in cloud-edge-IoT systems}
\label{tab:attack_defense_analysis}
\renewcommand{\arraystretch}{1.12}
\begin{tabular}{@{}p{0.13\textwidth}p{0.19\textwidth}p{0.13\textwidth}p{0.16\textwidth}p{0.30\textwidth}@{}}
\toprule
\textbf{Attack class} & \textbf{Threat assumptions and adversary capability} & \textbf{Goal and affected stage} & \textbf{Primary defenses} & \textbf{Defense limitations and tensions} \\
\midrule
Data/model poisoning and backdoors 
& Malicious clients or data owners can corrupt local data, labels, loss functions, or submitted updates; attackers may be single-client, colluding, or adaptive. 
& Availability degradation or targeted integrity failure during local training, update submission, and aggregation. 
& Robust aggregation, clipping, anomaly scoring, backdoor validation, client vetting, provenance. 
& Robust rules often assume concentrated benign updates; severe non-IID data can make honest minority updates appear anomalous. Secure aggregation can hide update-level signals needed for anomaly detection. DP noise and clipping may mask weak malicious signals or reduce the utility of rare classes. \\

Byzantine manipulation 
& A bounded or unbounded fraction of clients may submit arbitrary, coordinated, or optimization-aware updates independent of any local dataset. 
& Aggregation-time integrity and convergence disruption. 
& Median/trimmed mean, Krum, geometric median/RFA, sign-based aggregation, norm bounding. 
& Guarantees weaken under collusion, high dimensionality, adaptive attacks, client drift, and heterogeneous benign updates. Cryptographic confidentiality does not ensure the semantic correctness of encrypted updates. \\

Gradient leakage, membership inference, and property inference 
An honest-but-curious server, compromised aggregator, eavesdropper, or inference API user observes gradients, updates, model snapshots, or outputs. 
& Confidentiality breach during training, aggregation, or inference. 
& DP-SGD/PATE, secure aggregation, HE/MPC, TEEs, output control, privacy accounting. 
& Secure aggregation hides individual updates but not necessarily leakage from the final model. DP provides formal privacy but may reduce accuracy, fairness, and robustness, especially for small or non-IID clients. HE/MPC/TEEs add latency, memory, communication, energy, or hardware-trust costs. \\

Model extraction and IP theft 
& Querying adversary, malicious tenant, or insider can repeatedly access prediction APIs or deployed artifacts. 
& Model/IP theft during inference and deployment. 
& Rate limiting, confidence masking, query monitoring, watermarking, fingerprinting, and provenance. 
& Output restriction can reduce usability and calibration. Watermarking/fingerprinting supports ownership verification but does not prevent poisoning or privacy leakage and may be weakened by fine-tuning, pruning, or model distillation. \\

Network and orchestration attacks 
& On-path adversaries, compromised gateways, or orchestration faults can replay, delay, downgrade, substitute, or tamper with updates and control messages. 
& Confidentiality, integrity, and availability failures during communication, scheduling, and synchronization. 
& Authenticated transport, update signing, replay protection, attestation, secure aggregation, monitoring. 
& Channel security cannot stop malicious authenticated clients. Attestation and signing increase key management complexity and may not address denial-of-service attacks, stragglers, or stale-update attacks. \\

Sybil, free-rider, and game-theoretic attacks 
& Open or weakly vetted clients create multiple identities, submit low-effort updates, or strategically adapt to rewards, sampling, and detection rules. 
& Bias aggregation, manipulate incentives, or obtain benefits without useful contribution during admission, sampling, and training. 
& Identity management, capped sampling, reputation, contribution auditing, incentive design, anomaly monitoring. 
& Privacy-preserving aggregation limits direct auditing of contributions. Reputation systems may be gamed, and strong identity controls can be costly or impractical in cross-device IoT settings. \\
\bottomrule
\end{tabular}
\end{table*}

\subsection{Poisoning, Backdoors, and Byzantine Manipulation}
\label{sec:poisoning_byzantine}

Training-time attacks corrupt data, labels, local optimization, or communicated updates. In data poisoning, an adversary manipulates local training examples or labels to bias the learned decision function \cite{fung2018mitigating, yang2020adversary}. In model-update poisoning, the attacker directly submits malicious gradients or parameters that need not correspond to any real local dataset. These attacks may target availability by degrading convergence or accuracy or integrity by selectively changing model behavior for chosen inputs or classes.

Backdoor attacks are a structured form of integrity poisoning. The model behaves normally on clean inputs but misclassifies inputs containing an attacker-controlled trigger \cite{bagdasaryan2020backdoor, nguyen2024backdoor}. In FL, model-replacement backdoors are especially dangerous because a small number of malicious clients may implant persistent trigger behavior while preserving high clean accuracy.

Byzantine attacks generalize this threat model by allowing malicious clients to submit arbitrary or coordinated updates \cite{blanchard2017machine, yin2018byzantine}. Robust aggregation rules such as coordinate-wise median, trimmed mean, Krum, geometric median/Robust Federated Aggregation (RFA), and signSGD with majority vote can reduce the influence of corrupted updates. Still, their guarantees often weaken under strong non-IID data, collusion, or adaptive attack strategies \cite{pillutla2022robust, bernstein2018signsgd, li2023experimental}.

\subsection{Privacy Leakage, Model Extraction, and IP Threats}

DML can leak sensitive training information even when raw data are not explicitly shared. Gradient inversion and related reconstruction attacks can recover features or labels from shared gradients or updates, particularly when updates are fine-grained or insufficiently protected \cite{zhu2019deep, geiping2020inverting}. Membership inference attacks attempt to determine whether a specific record was used during training, with strong likelihood-ratio methods demonstrating substantial risk even under black-box access \cite{shokri2017membership, carlini2022membership}. Property inference can further reveal aggregate or sensitive attributes about a client population.

Model extraction attacks threaten intellectual property by using prediction APIs or exposed inference services to train surrogate models that approximate a proprietary model \cite{tramer2016stealing}. These attacks can enable competitive misuse, downstream adversarial transfer, or unauthorized redistribution. Watermarking, fingerprinting, output throttling, confidence masking, query monitoring, and extraction-aware API design are, therefore, important complements to privacy-preserving training.

\subsection{Network, Orchestration, and Sybil/Game-Theoretic Attacks}

DML repeatedly exchanges gradients, model updates, metadata, and control messages across networks. On-path adversaries may eavesdrop, replay, downgrade, substitute, or tamper with updates, enabling privacy leakage, model replacement, or poisoning in transit~\cite{bonawitz2017practical, zhu2019deep}. Authentication, encrypted transport, update signing, replay protection, secure aggregation, and round-level attestation are therefore necessary system-level protections.

Wireless and IoT deployments introduce additional physical-layer (PHY) and medium-access vulnerabilities because training depends on shared, time-varying communication channels \cite{angueira2022survey}. Jamming and selective interference can cause client dropout, delay synchronization, or bias participation toward clients with more reliable connectivity \cite{alhawas2026federated}. Spoofing, channel manipulation, and compromised gateways or access points can alter client identities, scheduling information, or transmitted updates \cite{butun2019security}. In over-the-air aggregation, where wireless signals are superimposed to compute an aggregate, adversarial manipulation of transmission power or waveform may directly bias the resulting update \cite{mendes2026survey}. Wireless backdoor and poisoning attacks can further exploit compromised edge devices and unreliable links while preserving apparently normal clean-model behavior~\cite{wan2024data}. Evaluations should therefore consider mobility, channel variation, link failures, selective jamming, and compromised network intermediaries alongside conventional poisoning and Byzantine attacks.

Open or weakly authenticated FL settings are also vulnerable to Sybil and free-rider attacks. Sybil adversaries create multiple fake clients to amplify poisoning, bias aggregation, or manipulate incentives~\cite{fung2018mitigating}. Free riders or lazy clients submit fabricated, noisy, or low-effort updates to obtain rewards without contributing useful information~\cite{fraboni2021free}. Game-theoretic models are useful for these settings because attackers and defenders adapt strategically through sampling policies, clipping thresholds, acceptance rules, incentives, and reputation mechanisms~\cite{bruckner2011stackelberg, chen2020learning, liu2024game}.

\subsection{How Heterogeneity Changes Attack Detectability and Defense Reliability}

Heterogeneity changes both attack feasibility and defense reliability. Under IID-like conditions, malicious updates may be easier to identify because benign updates cluster more tightly. Under realistic IoT heterogeneity, however, benign updates vary because of label skew, domain shift, device quality, sensor modality, and participation patterns. This dispersion can cause robust aggregation to reject useful minority-client updates or accept malicious updates that resemble benign outliers \cite{pillutla2022robust, li2023experimental, wang2024rfvir}.

Accordingly, attack evaluation should not rely on a single partition or clean aggregate metric. It should vary the number and participation frequency of malicious clients, the skew severity, the synchronization regime, the client sampling policy, and the defense configuration. Clean accuracy should be reported together with attack success rate (ASR), backdoor persistence, client-level disparity, privacy leakage, communication cost, latency, and energy.

\section{Defense Mechanisms and Tradeoffs}
\label{sec:defenses}

Securing DML requires coordinated defenses across the learning lifecycle. As summarized in Table~\ref{tab:defense_IoT_DML}, local-training defenses include drift-aware optimization and differential privacy. Before and during aggregation, secure computation and robust aggregation protect the confidentiality and integrity of updates. After aggregation, watermarking, fingerprinting, and provenance mechanisms support model ownership and accountability. These protections involve trade-offs: privacy mechanisms may reduce utility or increase latency, robust aggregation may reject valid minority-client updates, and cryptographic methods may increase computation, memory, and communication costs.

\begin{table}[h]
    \centering
    \caption{Defense mechanisms in IoT-oriented DML}
    \label{tab:defense_IoT_DML}
    \begin{tabular}{@{}p{1.5cm}cp{3.2cm}@{}}
        \toprule
        \textbf{Phase} & \textbf{Mechanism} & \textbf{Reference} \\
        \toprule
        \multirow{2}{*}{Local training} & Drift-aware optimization &
    \makecell[l]{~\cite{li2020federated, karimireddy2020scaffold, acar2021federated, reddi2020adaptive, li2021fedbn} ,\\
    ~\cite{fallah2020personalized, t2020personalized, li2021ditto} \\} 
        \\ \cmidrule(l){2-3} 
        & Differential privacy & \makecell[l]{~\cite{abadi2016deep, papernot2018scalable}} \\
        \cmidrule(l){2-3} 
        & Cryptographic protection &
        \makecell[l]{~\cite{cheon2017homomorphic, wu2025survey, yuan2025approximate},~\cite{mohassel2018aby3,knott2021crypten},~\cite{ohrimenko2016oblivious}} \\
        \midrule
        Before aggregation & Secure aggregation &
        ~\cite{bonawitz2017practical} \\
        \midrule
        During the aggregation & Robust aggregation &
        \cite{blanchard2017machine, yin2018byzantine, pillutla2022robust, bernstein2018signsgd} \\
        \midrule
        After aggregation & Watermarking &
        {~\cite{uchida2017embedding, adi2018turning}} \\
        \cmidrule(l){2-3} 
        & Fingerprinting & \cite{oliynyk2023know} \\
        \cmidrule(l){2-3} 
        & Provenance &
        {~\cite{nikolic2026model}} \\ \midrule 
        \bottomrule
    \end{tabular}
\end{table}

\subsection{Robust Aggregation and Drift-Aware Optimization}

Robust aggregation methods aim to limit the influence of faulty or malicious updates. Coordinate-wise median, trimmed mean, Krum, geometric-median-based RFA, and signSGD with majority vote are representative approaches \cite{blanchard2017machine, yin2018byzantine, pillutla2022robust, bernstein2018signsgd}. These methods can improve resilience to Byzantine behavior, but their effectiveness depends on the concentration of benign updates, the adversary fraction, dimensionality, and non-IID data.

Severe non-IID data can substantially degrade the performance of FL models protected by robust aggregation methods relative to IID settings \cite{li2023experimental}. Li et al.~\cite{li2023experimental} evaluated these methods under severe data heterogeneity using $\operatorname{Dirichlet}(\alpha=0.1)$. Under federated stochastic gradient descent (FedSGD), they reported poor accuracy with Krum, Geometric Median (GM), Auto-weighted Geometric Median (AutoGM) \cite{li2021byzantine}, and Median relative to their IID counterparts. Under Federated Averaging (FedAvg), performance similarly degraded across the evaluated aggregation methods.

The evaluated Byzantine attacks include inner-product manipulation (IPM), which crafts malicious updates to oppose the intended aggregate direction; sign flipping (SF), which reverses or negatively scales client updates; label flipping (LF), which corrupts local training labels; and A Little Is Enough (ALIE), which constructs malicious updates within the apparent statistical range of benign updates to evade detection.

On CIFAR-10 with FedAvg, Li et al.~\cite{li2023experimental} evaluated Centered Clipping (CC) with a fixed malicious-client fraction of 25\% and reported test accuracy under IID and non-IID data, using Dirichlet($\alpha=0.1$) for the non-IID partition. Alharbi et al.~\cite{alharbi2023robust} evaluated CC under non-IID data using Dirichlet($\alpha=0.5$), varied the number of attackers, and reported test error. Table~\ref{tab:model_poisoning_attack_analysis} summarizes the results of the two studies separately. Because the studies use different levels of heterogeneity, attacker configurations, and evaluation metrics, their numerical results are not directly comparable. Within each study, however, the results indicate that CC's performance can deteriorate substantially under poisoning attacks and non-IID data.

Non-IID data can alter both attack effectiveness and defense reliability. Backdoor attacks seek to associate inputs containing a trigger with an attacker-selected output. In the experiments of Zawad et al.~\cite{zawad2021curse}, some evaluated backdoor attacks became less effective as data heterogeneity increased. This result is limited to the experimental configurations examined in that study and should not be interpreted as evidence that non-IID data generally protects against backdoor attacks. Zawad et al.~\cite{zawad2021curse} also found that, in their experiments, non-IID training caused benign and malicious updates to exhibit similar overfitting characteristics, which reduced the effectiveness of defenses that used overfitting to identify malicious updates.

\begin{table*}[h]
\centering
\caption{Centered Clipping (CC) results reported by two independent studies on CIFAR-10 with FedAvg. The left column summarizes test accuracy from \cite{li2023experimental} using a fixed malicious-client fraction of 25\% under IID and non-IID data (Dirichlet $\alpha=0.1$ for the non-IID partition). The right column summarizes test error from \cite{alharbi2023robust} for non-IID data (Dirichlet, $\alpha=0.5$) as the number of attackers varies. Because the studies use different data partitions, attacker configurations, and evaluation metrics, their numerical results are not directly comparable}
\label{tab:model_poisoning_attack_analysis}
\renewcommand{\arraystretch}{1.12}
\begin{tabular}{p{3.5cm}p{6.5cm}p{6.5cm}}
\toprule
\textbf{Attack}
&
\textbf{Test accuracy reported in
\cite{li2023experimental}: IID and non-IID
($\alpha=0.1$), 25\% malicious clients}
&
\textbf{Test error reported in
\cite{alharbi2023robust}: non-IID
($\alpha=0.5$), varying attacker counts}
\\
\midrule

Inner Product Manipulation (IPM)
&
Under IPM with $\epsilon=0.5$, CC achieves 77.77\% test
accuracy with IID data and 56.97\% with non-IID data. Under
the stronger IPM attack with $\epsilon=100$, its test accuracy
falls to 10.00\% in both settings.
&
CC maintains a test error of approximately 60\% as the number
of IPM attackers increases from 3 to 18.
\\
\midrule

Sign Flipping (SF)
&
Under SF, CC achieves only 10.00\% test accuracy with both
IID and non-IID data.
&
CC reaches a test error of approximately 90\% when the number
of SF attackers exceeds 9.
\\
\midrule

Label Flipping (LF)
&
Under LF, CC achieves 69.80\% test accuracy with IID data and
52.41\% with non-IID data.
&
CC reaches a test error of approximately 90\% when the number
of LF attackers exceeds 12.
\\
\midrule

A Little Is Enough (ALIE)
&
Under ALIE, CC achieves 78.53\% test accuracy with IID data
and 59.29\% with non-IID data.
&
CC reaches a test error of approximately 90\% when the number
of ALIE attackers exceeds 12.
\\
\bottomrule
\end{tabular}
\end{table*}

Drift-aware optimization addresses the non-IID conditions that often undermine robustness. \textsc{FedProx} limits local drift through proximal regularization, \textsc{SCAFFOLD} uses control variates to correct client drift, \textsc{FedDyn} aligns local and global objectives, and \textsc{FedOpt} variants stabilize server-side optimization \cite{li2020federated, karimireddy2020scaffold, acar2021federated, reddi2020adaptive}. Modeling-layer methods such as \textsc{FedBN}, personalization, clustered FL, and client-specific heads can further reduce domain shift and fairness disparities \cite{li2021fedbn, fallah2020personalized, t2020personalized, li2021ditto}. In security-sensitive settings, these methods should be evaluated jointly with poisoning and Byzantine attacks, not only under benign non-IID training.

\subsection{Differential Privacy and Secure Aggregation}

Differential privacy protects individual training records by limiting the influence of any single example on the released model. Differentially private stochastic gradient descent (DP-SGD) clips per-example gradients. It injects calibrated noise, while Private Aggregation of Teacher Ensembles (PATE) uses noisy aggregation of teacher predictions to train a private student model \cite{abadi2016deep, papernot2018scalable}. These methods provide formal privacy guarantees but can reduce utility, especially for small clients, rare classes, or highly non-IID data~\cite{zhang2023trading}.

Secure aggregation protects communicated updates by ensuring that the server observes only the aggregate sum, not individual client updates \cite{bonawitz2017practical}. It is particularly useful under an honest-but-curious server model and is often combined with clipping and DP. However, secure aggregation does not, by itself, prevent poisoning or guarantee the privacy of the final model; it must be paired with robust aggregation, anomaly detection, and privacy accounting, tailored to the threat model.

\subsection{HE, MPC, and TEE-Based Protected Computation}

HE, MPC, and TEEs support protected computation when raw data, intermediate values, or local updates should not be exposed. HE enables selected computations over ciphertexts and is most mature for encrypted inference or HE-friendly linear operations. Still, it remains constrained by ciphertext expansion, expensive rotations, bootstrapping, and limited native support for nonlinear operators \cite{cheon2017homomorphic, wu2025survey, yuan2025approximate}. MPC allows parties to jointly compute functions without revealing private inputs, often using secret sharing or mixed protocols; systems such as ABY3 and CrypTen improve practicality but remain sensitive to communication, preprocessing, nonlinear operations, and network conditions \cite{mohassel2018aby3, knott2021crypten}. TEEs can execute richer ML logic with lower algorithmic redesign, but they rely on hardware trust, attestation, enclave memory limits, and side-channel resistance \cite{ohrimenko2016oblivious}.

Practically, HE is best suited to carefully redesigned encrypted inference or selected protected operations; MPC is attractive when participants can tolerate communication and preprocessing costs; and TEEs are often engineering-compatible when hardware trust is acceptable. Because energy reporting remains inconsistent, studies should report at least latency, communication volume, memory pressure, and, when possible, joules per inference or per training round.

\subsection{Watermarking, Fingerprinting, and Provenance}

Model watermarking and fingerprinting protect DML model intellectual property by embedding verifiable ownership signals into parameters or model behavior. White-box methods may embed signatures directly into weights, whereas black-box methods use trigger or behavior-based verification through the prediction interface \cite{tekgul2021waffle, uchida2017embedding, adi2018turning, oliynyk2023know}. Effective ownership mechanisms should preserve utility, remain stealthy, survive compression or fine-tuning, and support defensible post hoc verification.

Provenance mechanisms complement watermarking. Signed model artifacts, versioned checkpoints, client-update logs, secure audit trails, and deployment metadata help establish which data, clients, aggregation rules, and defenses contributed to a released model~\cite{nikolic2026model}. In regulated IoT settings, provenance also supports accountability, rollback, incident response, and evidence of compliance.

\subsection{Defense Composition: Non-Compositional Trade-offs}
\label{sec:defense_composition}

No defense provides comprehensive protection, and combining defenses does not necessarily yield additive benefits. Secure aggregation improves confidentiality by concealing individual client updates, but this reduced visibility can limit anomaly detection, client scoring, and robust aggregation~\cite{bonawitz2017practical, kairouz2021advances, pillutla2022robust}. Similarly, DP-induced clipping and noise may reduce the statistical separation between benign heterogeneous updates and weak poisoning or backdoor signals. At the same time, aggressive robust filtering may reject useful updates from honest minority clients~\cite{abadi2016deep, zhang2023trading, chen2023privacy}.

The threat model should therefore guide defense composition, required update visibility, heterogeneity level, resource constraints, and residual trust assumptions. Compatibility should be evaluated jointly across confidentiality, integrity, privacy, utility, and operational cost rather than inferred from the effectiveness of each defense in isolation. Section~VI and Table~\ref{tab:dml_usecases} provide deployment-specific guidance and reporting criteria for selecting and evaluating such defense portfolios.

\section{Deployment, Evaluation, and Governance Guidelines}
\label{sec:deployment_guidelines}

Deployment context determines which threats are most relevant, which defenses are practical, and which evaluation metrics should be reported. In IoT-oriented DML, security design is shaped not only by data sensitivity and trust assumptions, but also by device constraints, intermittent connectivity, heterogeneous sensing modalities, physical exposure, gateway reliability, and edge-cloud orchestration. Representative deployments include industrial sensing, vehicular IoT, smart healthcare and wearables, smart cities/cyber-physical system (CPS), IoT cybersecurity, and cloud-edge IoT services. Table~\ref{tab:dml_usecases} summarizes these settings by mapping each domain to its typical topology, IoT-specific constraints and threats, recommended defense practices, and reporting metrics.

\begin{table*}[h]
\centering
\tiny
\setlength{\tabcolsep}{3pt}
\renewcommand{\arraystretch}{1.12}
\caption{IoT-centered DML deployment settings, constraints, design guidelines, and reporting metrics. \textit{Abbrev.:} DP = differential privacy; DP-SGD = differentially private stochastic gradient descent; HE = homomorphic encryption; MPC = secure multiparty computation; TEE = trusted execution environment; RSU = roadside unit; ASR = attack success rate; SLO = service-level objective; OEE = overall equipment effectiveness; PHI = protected health information; AUROC = area under the receiver operating characteristic curve; AUPRC = area under the precision-recall curve; DDoS = distributed denial of service; IDS = intrusion detection system; TPR = true-positive rate; FPR = false-positive rate}
\label{tab:dml_usecases}
\begin{tabular}{p{2cm}|p{2.2cm}|p{2.8cm}|p{3.25cm}|p{3.35cm}|p{2.9cm}}
\hline
\textbf{IoT domain / setting} & \textbf{Use cases} & \textbf{DML pattern / topology} & \textbf{IoT-specific constraints and threats} & \textbf{Recommended defenses / practices} & \textbf{Metrics to report} \\
\hline

Industrial sensing / IIoT 
& Predictive maintenance; quality inspection; adaptive scheduling 
& Cross-silo FL; plant gateway-edge/cloud hybrid; periodic or event-driven synchronization 
& Plant/site skew, heterogeneous sensors, legacy devices, intermittent links, constrained gateways, physical tampering, gateway compromise, data locality, and operational safety 
& Authenticated transport, device identity management, gateway-level anomaly screening, non-IID-aware aggregation, robust aggregation, provenance, drift monitoring, and fail-safe rollback policies \cite{shafique2018overview,penumuru2019identification,dalzochio_machine_2020,khan_industrial_2020,younan_challenges_2020} 
& OEE/throughput, defect rate, downtime, false alarms, ASR, communication, latency, memory, energy, and fairness across lines/sites \\
\hline

Vehicular IoT 
& Cooperative perception; fleet learning; controller/model updates 
& Cross-device FL; hierarchical vehicle-RSU-edge/cloud learning 
& High mobility, spatiotemporal skew, intermittent V2X links, stale updates, strict tail latency, physical/environmental interference, and safety-critical poisoning/backdoors 
& Hierarchical aggregation, stale-update control, signed updates, distribution-aware robust aggregation, poisoning/backdoor stress tests, edge placement for latency-sensitive tasks, and local fallback policies \cite{zeng2021federated,khan2021federated,li2020federated,yang2020adversary} 
& mAP or task utility, tail/edge latency, stale-update rate, link dropout, ASR, backdoor persistence, client-drift metrics, and communication/energy cost \\
\hline

Smart healthcare and wearables 
& Wearable monitoring; collaborative diagnostics; outcome prediction; imaging 
& Cross-device wearable FL; cross-silo hospital FL; edge/cloud orchestration 
& PHI sensitivity, patient/site-level non-IID data, cohort shift, label imbalance, battery limits, small memory, intermittent participation, sensor drift, missing measurements, and device loss or compromise 
& Lightweight local training, dropout-tolerant secure aggregation, DP\text{-}SGD/PATE with explicit privacy accounting, selected HE/MPC or TEE support for sensitive operations, gateway buffering, attestation for managed devices, and fairness auditing across patients/sites \cite{teo2024federated,qayyum2021collaborative,li2020federated,khan2021federated} 
& AUROC/AUPRC, DP $\epsilon/\delta$, per-site or per-patient disparity, dropout rate, ASR, latency, memory, communication, and battery/energy overhead \\
\hline

Smart cities / CPS 
& Traffic control; energy response; environmental monitoring; anomaly detection 
& Cross-domain FL across agencies/utilities; edge federation across cameras, meters, roadside units, and gateways 
& Device heterogeneity, district-level domain shift, heterogeneous ownership, PII exposure, public-service reliability, physical-layer disruption, and unreliable clients/gateways 
& Cross-domain governance, secure aggregation, DP when PII is present, personalization/FedBN or clustered FL for spatial skew, authenticated communication, audit logs, and resilience testing under device, gateway, and link failures \cite{ahmad2020developing,liu2018cps,qian_towards_2018,gupta_smart_2020,khan2021federated} 
& Service reliability, availability, travel-time or response utility, privacy budget, ASR, fairness by district/site, communication overhead, latency, and energy \\
\hline

IoT cybersecurity 
& DDoS/IDS; malware detection; log and telemetry analytics 
& Cross-tenant or cross-organization FL; streaming edge telemetry; gateway/cloud aggregation 
& Highly non-IID traffic, adaptive adversaries, poisoned telemetry, compromised clients/gateways, tenant isolation, model/IP exposure, and evolving attacks 
& Byzantine-robust aggregation, clipping/norm bounds, secure aggregation where compatible, attack-aware benchmarks, provenance, authenticated telemetry, query/interface monitoring, and model watermarking/fingerprinting when IP risk is high \cite{yang2020adversary,rodriguez2023survey,qayyum2020securing,yang2023PrivacyandModelIP} 
& TPR@FPR, false positives, ASR/backdoor persistence, detection latency, communication, memory, energy, and model-extraction risk \\
\hline

Cloud-edge IoT services 
& Managed IoT training/inference; federated MLOps; edge analytics 
& Device/gateway streams with cloud-hosted training, inference, monitoring, and rollback 
& Edge-cloud placement, multi-tenancy, access control, model extraction, orchestration faults, data residency, cost, and energy/carbon constraints 
& Access control, authenticated APIs, rate limiting, output control, provenance, deployment gates, model rollback, explicit placement of filtering/aggregation/privacy accounting, and energy/carbon-aware scheduling \cite{sitton-candanedo_review_2019,qayyum2020securing,yousefpour2023green,thakur2025green} 
& Cost/hour, time-to-accuracy, SLO adherence, tail latency, extraction risk, memory, energy, kgCO$_2$e, and rollback frequency \\
\hline

\end{tabular}
\end{table*}

\subsection{IoT-Domain Deployment Constraints}

Industrial IoT and smart factories require DML designs that tolerate plant-level skew, heterogeneous sensors, legacy devices, intermittent links, gateway faults, and physical manipulation. In these settings, model failures can affect product quality, safety, downtime, and operational continuity; therefore, defenses should combine authenticated communication, device identity management, gateway-level anomaly screening, drift monitoring, robust aggregation, provenance, and fail-safe rollback policies \cite{shafique2018overview, penumuru2019identification, dalzochio_machine_2020}. Vehicular IoT and connected/autonomous vehicles operate under a different regime: highly mobile clients, intermittent vehicle-to-everything (V2X) connectivity, spatiotemporal data skew, strict tail-latency requirements, and safety-critical exposure to poisoning, backdoors, stale updates, sensor spoofing, and jamming. These deployments, therefore, require hierarchical vehicle-RSU-edge/cloud aggregation, stale-update control, signed updates, distribution-aware robustness, backdoor stress testing, and local fallback policies when communication or aggregation is delayed \cite{zeng2021federated, li2020federated, yang2020adversary}.

Smart healthcare and wearable systems introduce privacy-sensitive, patient-level non-IID data collected from battery-powered, memory-constrained, and intermittently connected devices. Wearable clients may also experience sensor drift, missing measurements, device loss, or compromise. As a result, IoT-health deployments should favor lightweight local training, dropout-tolerant secure aggregation, carefully budgeted DP, fairness auditing across patients or clinical sites, and stronger attestation or audit controls for managed medical gateways \cite{qayyum2021collaborative, teo2024federated}. 

Human-centered IoT sensing also includes behavioral biometrics, in which user-specific patterns captured by commodity devices support authentication and personalization \cite{essahraui2026comprehensive}. For example, Huang et al.~\cite{huang2026identifying} developed a smartphone-based, content-independent handwriting authentication system that abstracts identity-discriminative writing style from the specific written content. Such applications illustrate an important trustworthy-learning tension: behavioral signals must remain sufficiently stable for identity recognition while being robust to content variation, sensor variability, imitation, and distribution shift \cite{liang2020behavioral}. When these models are trained collaboratively, raw handwriting trajectories, learned representations, and shared updates should also be treated as identity-sensitive information requiring privacy protection, secure aggregation, and robustness evaluation.

Smart-city and CPS deployments combine heterogeneous ownership, district-level domain shift, public-service reliability requirements, personally identifiable data, and physical-layer disruptions affecting cameras, meters, roadside units, or environmental sensors. These settings require cross-domain governance, secure aggregation, DP when personally identifiable information (PII) is present, personalization or clustered FL for spatial skew, audit logs, and resilience testing under device, gateway, and link failures \cite{ahmad2020developing,liu2018cps}.

Cloud-edge IoT services add orchestration-level constraints that generic FL threat models do not capture. Training, filtering, aggregation, monitoring, privacy accounting, and rollback may be split across devices, gateways, edge servers, and cloud platforms. This placement affects latency, energy, communication cost, data residency, attack visibility, and trust boundaries. Cloud-hosted DML and MLaaS further introduce multi-tenancy, access control, model extraction, API abuse, cost, and carbon constraints, motivating authenticated interfaces, rate limiting, output control, provenance, deployment gates, model rollback, and explicit reporting of energy and carbon overhead \cite{qayyum2020securing, verbraeken2020survey, googleTPU,aws, microsoftAzure, ibmCloud, liu2025data}.

Across these domains, security claims should be scoped to the concrete IoT topology rather than stated generically. Cross-silo healthcare or industrial FL may tolerate stronger identity management, TEEs, MPC, and audit trails. In contrast, cross-device wearable or vehicular FL often requires lightweight client sampling, asynchronous participation, dropout-tolerant aggregation, energy-aware scheduling, and fallback operation. Thus, trustworthy IoT-oriented DML should report not only clean utility and attack success rate, but also client dropout, stale-update rate, communication overhead, memory footprint, tail latency, energy consumption, fairness across sites or device groups, and residual trust assumptions.

\subsection{Attack-Aware Benchmarks and Reproducibility}

DML evaluations should be reproducible and attack-aware. For non-IID studies, authors should specify the skew type, partition-generation protocol, Dirichlet $\alpha$ or equivalent parameter, client sampling policy, number of clients, malicious-client fraction, and random seeds. When possible, studies should use open-source evaluation frameworks or release partitions and scripts \cite{caldas2018leaf, chai2024survey}.

For security evaluation, attack settings should include poisoning, backdoors, Byzantine manipulation, gradient leakage, membership inference, model extraction, and network tampering when relevant to the claimed threat model. Robustness should be tested under multiple skew severities rather than a single near-IID partition. Backdoor studies should report both clean accuracy and ASR; privacy studies should report $\epsilon/\delta$, attack success, and any utility loss; and robust aggregation studies should report breakdown behavior under different malicious-client fractions and non-IID regimes.

\subsection{Non-IID Data Partition}

Dirichlet partitioning is one of the most widely used approaches for simulating non-IID data~\cite{li2023experimental, thein2024personalized, jimenez2025thorough,zeng2024bsr}. Its concentration parameter $\alpha$ controls the degree of heterogeneity: smaller values, such as $0.1$-$0.5$, generally produce highly skewed client distributions, whereas larger values produce more nearly uniform distributions~\cite{li2023experimental}. Another common approach is class-based partitioning, which restricts the number of classes assigned to each client. For example, the one-class and two-class settings in~\cite{zhao2018federated} assign data from only one or two classes, respectively, to each client. To model heterogeneity that varies across both geographic locations and time, Jimenez et al.~\cite{jimenez2025thorough} employed the St-Dirichlet method to generate spatiotemporal non-IID partitions. Alternatively, benchmark suites such as LEAF~\cite{caldas2018leaf} and OARF~\cite{hu2022oarf} provide naturally heterogeneous federated datasets for more realistic evaluation.

\subsection{Non-IID Metrics}

To determine whether a data distribution is non-IID, a quantitative metric is often used to measure its distance from a global (aggregated) distribution. As research on quantifying the level of non-IID advances, common distribution-based non-IID metrics, including EMD ~\cite{zhao2018federated}, Hellinger distance, Jensen-Shannon divergence~\cite{lai2022fedscale}, and KL divergence~\cite{le2025fedklentropy}, are being used to characterize the non-IID in label distribution, feature distributions, and data volume distributions across participating clients~\cite{Gutierrez2024NonIID}. For example, during the FL initialization phase, EMD was used to measure the total cost of moving a client's label distribution to another client 's~\cite {zhao2018federated}. A large cost value indicates a significant difference between the two local distributions~\cite{zhao2018federated}, leading to poor performance of the FL global model. Furthermore, the Jensen-Shannon divergence quantifies the dissimilarity between pairs of client label distributions, revealing the extent of distributional heterogeneity across federated image, audio, video, and text datasets~\cite{lai2022fedscale}. In the FL model aggregation phase, the KL divergence allows the server to adaptively assign weights based on how well a client's model weight distribution aligns with the aggregated model weight distribution~\cite{le2025fedklentropy}.

\subsection{Metrics Beyond Accuracy: ASR, Privacy Budget, Disparity, Latency, Communication, and Energy}

Average test accuracy is insufficient for trustworthy DML. At minimum, evaluation should report clean model utility, attack success rate~\cite{sanchez2024federatedtrust}, poisoned-versus-clean trade-offs, privacy budget~\cite{tanveer2025balancing, wagner2018technical}, client-level or site-level disparity, convergence behavior~\cite{koloskova2022sharper}, communication overhead~\cite{kim2024spafl}, latency~\cite{thapa2022splitfed}, memory~\cite{tanveer2025balancing}, and energy~\cite{xia2024towards}. For safety-critical or IoT settings, additional metrics may include tail latency, availability, rollback frequency, fairness by site or district, domain-drift indicators, and service-level objective adherence~\cite{khan2021federated, li2024holistic}.

Energy and carbon should be reported using a clear boundary, such as client-side compute, server-side compute, enclave, or cryptographic accelerator costs, and network transfer energy. When direct power measurement is unavailable, latency and communication volume should be reported as first-order proxies, especially for HE-, MPC-, TEE-, and secure-aggregation-based systems.

\subsection{Open-Source Evaluation Frameworks}
Evaluation-oriented techniques aim to assess and compare FL algorithms according to diverse dimensions: non-IID data partitions (data heterogeneity) and device heterogeneity~\cite{zhang2024flhetbench}, data-type and task scalability~\cite{lai2022fedscale}, robustness against adversarial threats~\cite{gonzalez2025byzflresearchframeworkrobust, zhang2025flpoison, han2024fedsecurity}. Each offers structured guidance for implementing new algorithms, along with standardized evaluation protocols for fair and replicable comparisons. Beyond test accuracy metrics, FedScale~\cite{lai2022fedscale} provides software tools to measure system performance in terms of wall-clock time, communication cost, and device availability duration~\cite{lai2022fedscale}. 

LEAF~\cite{caldas2018leaf} provides realistic FL datasets while~\cite{li2022federated} simulated non-IID data silo cases by using six different partition strategies. Under the effects of data heterogeneity, BYZFL~\cite{gonzalez2025byzflresearchframeworkrobust}, FedSecurity~\cite{han2024fedsecurity}, and FLPoison~\cite{zhang2025flpoison} focus on systematically evaluating the robustness of various defense mechanisms against data, model poisoning, and FL privacy attacks using ASR and other specific metrics. Table~\ref{tab:FL_benchmarks} illustrates a comparative analysis of the open-source FL algorithm benchmarking tools.

FLHetBench~\cite{zhang2024flhetbench} evaluates FL algorithms under differences in device computational capability and availability. Using this framework, Zhang et al. found that longer wall-clock times and lower resource utilization were associated with lower model accuracy in the device-heterogeneous settings they evaluated. Other benchmarking frameworks provide standardized evaluation protocols and metrics for multimodal data~\cite{feng2023fedmultimodal}, large language models~\cite{han2024fedsecurity}, and safety-critical domains~\cite{ogier2022flamby}.

\begin{table*}[h]
\centering
\caption{Comparison of open-source federated learning benchmark tools}
\label{tab:FL_benchmarks}
\renewcommand{\arraystretch}{1.12}
\begin{tabular}{p{1.0cm}p{4.9cm}p{4.9cm}p{4.9cm}}
\toprule
\textbf{Ref.} &
\textbf{Non-IID data partitioning} &
\textbf{Supported attacks and defenses} &
\textbf{Reported metrics} \\
\midrule

\cite{caldas2018leaf}
&
Federated image datasets (FEMNIST and CelebA), text datasets
(Sentiment140, Shakespeare, and Reddit), and a synthetic tabular
dataset.
&
No support.
&
Statistical metrics: device-level test accuracy, training loss, and
number of samples per user. System metrics: total computation across
devices, measured in FLOPs, and client-server network traffic,
measured in bytes written.
\\
\midrule

\cite{li2022federated}
&
Label-distribution skew based on fixed-sample and Dirichlet
partitioning; feature-distribution skew based on Gaussian and
cube-based sampling; quantity skew based on a Dirichlet distribution;
and FEMNIST.
&
No support.
&
Test accuracy, computation time (seconds), and communication volume
(MB).
\\
\midrule

\cite{lai2022fedscale}
&
Twenty real-world FL datasets spanning different scales and task
categories, together with a client system-behavior dataset.
&
Limited support.
&
Accuracy or loss, rounds to target accuracy, time to target accuracy,
power (W), energy (J), latency (seconds), wall-clock time,
communication cost, and device-availability duration.
\\
\midrule

\cite{gonzalez2025byzflresearchframeworkrobust}
&
Dirichlet- and gamma-similarity-based distributions.
&
\textbf{Attacks:} six Byzantine attacks.
\textbf{Defenses:} four pre-aggregation methods and ten
Byzantine-robust aggregation methods.
&
Test accuracy, training loss, and worst-case accuracy.
\\
\midrule

\cite{han2024fedsecurity}, \cite{he2020fedml}
&
Dirichlet partitioning; LEAF datasets (FEMNIST and Shakespeare);
and PubMedQA.
&
\textbf{Attacks:} four model-poisoning, two data-poisoning, and three
data-reconstruction methods.
\textbf{Defenses:} twelve pre-aggregation, three aggregation-stage,
and two post-aggregation methods.
&
Test accuracy and test loss.
\\
\midrule

\cite{zhang2025flpoison}
&
Dirichlet partitioning and FEMNIST.
&
\textbf{Attacks:} eight model-poisoning, two data-poisoning, and five
hybrid-poisoning methods.
\textbf{Defenses:} six methods against data poisoning and fifteen
methods against model poisoning.
&
Test accuracy, attack success rate (ASR), targeted attack impact, and
targeted defense robustness.
\\

\bottomrule
\end{tabular}
\end{table*}

\subsection{Governance Artifacts and Compliance-Oriented Reporting}

Governance translates technical guarantees into deployable evidence. A trustworthy DML deployment should document the threat model, trust assumptions, client admission policy, aggregation rule, privacy mechanism, non-IID evaluation protocol, attack suite, known defense limitations, and rollback or incident-response procedure \cite{chen2024trustworthy}. Audit logs, signed checkpoints, model cards, data-use records, privacy-accounting reports, and provenance metadata support both internal accountability and external compliance.

Security should also be integrated into MLOps \cite{patel2025towards}. Promotion gates should require attack-aware validation before a model is deployed; runtime monitors should track drift, abnormal client behavior, ASR indicators, and privacy-relevant query patterns; and rollback mechanisms should be available when poisoning, backdoors, or severe performance drift are detected. In this view, adversarial robustness is not a one-time experiment but a lifecycle property that must be monitored, documented, and periodically revalidated \cite{carlini2019evaluating}.

\section{Open Research Directions}
\label{sec:open}

The preceding sections show that trustworthy DML in cloud-edge-IoT systems cannot be achieved through isolated defenses. Robustness, privacy, heterogeneity management, protected computation, incentive design, reproducibility, and governance must be treated as interacting design requirements. This section summarizes the main open research directions that follow from this security-centered view.

\subsection{Heterogeneity-Aware Security Benchmarks}

A major barrier to progress is the lack of widely accepted benchmarks for evaluating attacks and defenses under realistic non-IID conditions. Existing studies often use different partition generators, skew severities, client-sampling policies, attacker fractions, and reporting metrics, making results difficult to compare. Future benchmarks should provide public partitions, fixed seeds, explicit heterogeneity diagnostics, and standardized attack suites covering poisoning, backdoors, Byzantine behavior, privacy leakage, and model extraction. Metrics such as Dirichlet $\alpha$, Earth Mover's Distance, Hellinger distance, Jensen-Shannon divergence, quantity-skew indices, and client-level performance should be reported alongside clean accuracy and attack success rate \cite{chai2024survey, jimenez2025thorough, nguyen2024backdoor}.

\subsection{Distribution-Aware Robustness Under Adaptive Attacks}

Robust aggregation remains fragile when benign updates are naturally dispersed due to label skew, feature shift, class imbalance, or spatiotemporal variation. Future work should develop defenses whose breakdown behavior is explicitly characterized under heterogeneous benign updates rather than near-IID assumptions. This includes robust aggregation, client vetting, anomaly scoring, and backdoor detection methods that distinguish malicious deviations from legitimate minority-client behavior. Evaluation should vary skew severity, malicious-client fraction, attacker participation frequency, synchronization regime, and client-sampling policy, while reporting both global and per-client outcomes \cite{pillutla2022robust, li2023experimental, jimenez2025thorough}.

\subsection{End-to-End Privacy Leakage Measurement}

Privacy leakage in DML depends on training mode, architecture, normalization layers, gradient granularity, clipping, mixed precision, microbatching, and whether secure aggregation is enabled. Yet privacy evaluations are often conducted under simplified configurations. Future studies should jointly report formal privacy budgets, reconstruction quality, membership inference performance, model utility, and system overhead under realistic training pipelines. DP-SGD, PATE, secure aggregation, output control, and cryptographic protection should be evaluated as composable defenses rather than isolated mechanisms \cite{abadi2016deep, papernot2018scalable, bonawitz2017practical, carletti2025sok, leite2024federated}.

\subsection{Practical Protected Computation for IoT DML}

HE, MPC, and TEEs are increasingly relevant for privacy-preserving DML, but their deployment costs remain highly workload-dependent. HE still faces ciphertext expansion, bootstrapping cost, rotation overhead, and limited native support for nonlinear operators. MPC can support richer computation but often shifts cost to communication, preprocessing, and interaction. TEEs preserve more standard ML execution patterns but rely on hardware trust, attestation, enclave-memory limits, and side-channel resistance. Future research should prioritize predictable performance, compiler support, operator libraries, autotuning, profiling, and end-to-end integration with DML workflows, especially for latency- and energy-constrained IoT deployments \cite{cheon2017homomorphic, mohassel2018aby3, knott2021crypten, ohrimenko2016oblivious, wu2025survey, yuan2025approximate}.

\subsection{Joint Robustness, Privacy, Fairness, and Efficiency Optimization}

Many defenses improve one dimension while weakening another. Differential privacy can reduce leakage but may degrade utility or exacerbate client-level disparities in severely non-IID data. Secure aggregation protects individual updates but can limit server-side anomaly detection. Robust aggregation can reduce the impact of poisoning but may suppress valid minority-client updates. Cryptographic protection can strengthen confidentiality but increase latency, memory, communication, and energy costs. Future work should move toward Pareto-style evaluation that jointly reports clean utility, ASR, privacy budget, per-client disparity, convergence, latency, communication, memory, energy, and carbon footprint \cite{lai2022fedscale, yousefpour2023green, thakur2025green}.

\subsection{Incentive-Compatible and Sybil-Resistant Participation}

Open and cross-device DML systems are vulnerable to free-riding, sparse participation, collusion, and Sybil behavior. Existing defenses often treat malicious clients as static adversaries, whereas real participants may adapt strategically to rewards, privacy costs, computational burdens, and detection policies. Future research should integrate mechanism design, reputation, attestation, contribution valuation, and participation control to encourage truthful behavior while limiting adversarial influence. Such methods must remain effective under non-IID data, intermittent availability, and privacy-preserving aggregation, where direct inspection of local updates may be limited \cite{fung2018mitigating, fang2020local, fraboni2021free, tan2024long}.

\subsection{Governance, Reproducibility, and Lifecycle Assurance}

Secure DML research also needs stronger reproducibility and governance practices. Reproducible studies should release partitions, seeds, attack scripts, defense configurations, telemetry, and system-cost measurements where possible. Governance-oriented deployments should translate technical controls into auditable artifacts, including threat models, privacy-accounting reports, model cards, data-use records, signed checkpoints, risk registers, and incident-response procedures. Future work should map technical guarantees to regulatory and institutional requirements so that secure aggregation, DP, provenance, and monitoring can support compliance-oriented review rather than remain purely technical claims \cite{lai2022fedscale, chai2024survey, smuha2025regulation, ai2023artificial}.

Overall, future progress in trustworthy DML will depend on integrated, deployment-aware research that jointly addresses heterogeneity, privacy leakage, adversarial robustness, protected computation, incentive alignment, reproducibility, and governance. The field should move beyond isolated algorithmic defenses toward lifecycle-oriented secure DML systems that can be evaluated, deployed, monitored, and audited under realistic cloud-edge-IoT conditions.

\section{Conclusion}
\label{sec:conclusion}

Distributed machine learning (DML) enables collaborative model development without requiring complete data centralization, but it also introduces interconnected risks to security, privacy, reliability, and fairness. This survey examined these risks across conventional distributed learning and federated learning in cloud-edge-IoT environments.

A central conclusion is that heterogeneity is not merely an optimization issue. Non-IID data, resource asymmetry, intermittent participation, and unequal trust can affect attack effectiveness, anomaly detection, robust aggregation, privacy protection, and client-level fairness. Security evaluations must therefore state their threat and trust assumptions and test defenses under representative heterogeneity and system conditions.

No individual mechanism protects the entire DML lifecycle. Dependable deployments require complementary controls for local training, aggregation, communication, model ownership, monitoring, and recovery. Their evaluation should extend beyond clean accuracy to include attack effectiveness, privacy guarantees, client-level disparities, system overhead, and reproducibility. Progress toward trustworthy DML will depend on defenses that remain effective, auditable, and operationally practical under realistic cloud-edge-IoT conditions.

\bibliographystyle{IEEEtran}
\bibliography{refs.bib}

@inproceedings{mcmahan2017communication,
  title={Communication-efficient learning of deep networks from decentralized data},
  author={McMahan, Brendan and Moore, Eider and Ramage, Daniel and Hampson, Seth and y Arcas, Blaise Aguera},
  booktitle={Artificial intelligence and statistics},
  pages={1273--1282},
  year={2017},
  organization={Pmlr}
}

@inproceedings{allouah2023fixing,
  title={Fixing by mixing: A recipe for optimal byzantine ml under heterogeneity},
  author={Allouah, Youssef and Farhadkhani, Sadegh and Guerraoui, Rachid and Gupta, Nirupam and Pinot, Rafa{\"e}l and Stephan, John},
  booktitle={International Conference on Artificial Intelligence and Statistics},
  pages={1232--1300},
  year={2023},
  organization={PMLR}
}

@inproceedings{karimireddy2020byzantine,
  title     = {Byzantine-Robust Learning on Heterogeneous Datasets via Bucketing},
  author    = {Karimireddy, Sai Praneeth and He, Lie and Jaggi, Martin},
  booktitle = {International Conference on Learning Representations},
  year      = {2022},
  url       = {https://openreview.net/forum?id=jXKKDEi5vJt}
}

@article{yang2020adversary,
  title={Adversary-resilient distributed and decentralized statistical inference and machine learning: An overview of recent advances under the Byzantine threat model},
  author={Yang, Zhixiong and Gang, Arpita and Bajwa, Waheed U},
  journal={IEEE Signal Processing Magazine},
  volume={37},
  number={3},
  pages={146--159},
  year={2020},
  publisher={IEEE}
}

@article{gonzalez2025byzflresearchframeworkrobust,
  title={Byzfl: Research framework for robust federated learning},
  author={Gonz{\'a}lez, Marc and Guerraoui, Rachid and Pinot, Rafael and Rizk, Geovani and Stephan, John and Ta{\"\i}ani, Fran{\c{c}}ois},
  journal={arXiv preprint arXiv:2505.24802},
  year={2025}
}

@inproceedings{186214,
    title = {{Scaling Distributed Machine Learning with the Parameter Server}},
    year = {2014},
    booktitle = {11th {\{}USENIX{\}} Symposium on Operating Systems Design and Implementation ({\{}OSDI{\}} 14)},
    author = {Li, Mu and Andersen, David G and Park, Jun Woo and Smola, Alexander J and Ahmed, Amr and Josifovski, Vanja and Long, James and Shekita, Eugene J and Su, Bor-Yiing},
    month = {10},
    pages = {583--598},
    publisher = {{\{}USENIX{\}} Association},
    url = {https://www.usenix.org/conference/osdi14/technical-sessions/presentation/li_mu},
    address = {Broomfield, CO},
    isbn = {978-1-931971-16-4}
}

@misc{NVIDIADeveloper,
    title = {{NVIDIA Collective Communications Library (NCCL) | NVIDIA Developer}},
    url = {https://developer.nvidia.com/nccl}
}

@article{mayer2020scalable,
  title={Scalable deep learning on distributed infrastructures: Challenges, techniques, and tools},
  author={Mayer, Ruben and Jacobsen, Hans-Arno},
  journal={ACM Computing Surveys (CSUR)},
  volume={53},
  number={1},
  pages={1--37},
  year={2020},
  publisher={ACM New York, NY, USA}
}

@article{Yuan2019AdversarialLearning,
    title = {{Adversarial Examples: Attacks and Defenses for Deep Learning}},
    year = {2019},
    journal = {IEEE transactions on neural networks and learning systems},
    author = {Yuan, Xiaoyong and He, Pan and Zhu, Qile and Li, Xiaolin},
    number = {9},
    month = {9},
    pages = {2805--2824},
    volume = {30},
    publisher = {NLM (Medline)},
    doi = {10.1109/TNNLS.2018.2886017},
    issn = {21622388},
    pmid = {30640631},
    arxivId = {1712.07107}
}

@inproceedings{shafique2018overview,
  title={An overview of next-generation architectures for machine learning: Roadmap, opportunities and challenges in the IoT era},
  author={Shafique, Muhammad and Theocharides, Theocharis and Bouganis, Christos-Savvas and Hanif, Muhammad Abdullah and Khalid, Faiq and Haf{\i}z, Rehan and Rehman, Semeen},
  booktitle={2018 Design, Automation \& Test in Europe Conference \& Exhibition (DATE)},
  pages={827--832},
  year={2018},
  organization={IEEE}
}

@article{verbraeken2020survey,
  title={A survey on distributed machine learning},
  author={Verbraeken, Joost and Wolting, Matthijs and Katzy, Jonathan and Kloppenburg, Jeroen and Verbelen, Tim and Rellermeyer, Jan S},
  journal={ACM Computing Surveys (CSUR)},
  volume={53},
  number={2},
  pages={1--33},
  year={2020},
  publisher={ACM New York, NY, USA}
}

@article{qayyum2020securing,
  title={Securing Machine Learning in the Cloud: A Systematic Review of Cloud Machine Learning Security},
  author={Qayyum, Adnan and Ijaz, Aneeqa and Usama, Muhammad and Iqbal, Waleed and Qadir, Junaid and Elkhatib, Yehia and Al-Fuqaha, Ala},
  journal={Frontiers in big Data},
  volume={3},
  year={2020},
  publisher={Frontiers Media SA}
}

@article{ahmad2020developing,
  title={Developing Future Human-Centered Smart Cities: Critical Analysis of Smart City Security, Interpretability, and Ethical Challenges},
  author={Ahmad, Kashif and Maabreh, Majdi and Ghaly, Mohamed and Khan, Khalil and Qadir, Junaid and Al-Fuqaha, Ala},
  journal={arXiv preprint arXiv:2012.09110},
  year={2020}
}

@article{li2020federated,
  title={Federated learning: Challenges, methods, and future directions},
  author={Li, Tian and Sahu, Anit Kumar and Talwalkar, Ameet and Smith, Virginia},
  journal={IEEE Signal Processing Magazine},
  volume={37},
  number={3},
  pages={50--60},
  year={2020},
  publisher={IEEE}
}

@article{zeng2021federated,
  title={Federated Learning on the Road: Autonomous Controller Design for Connected and Autonomous Vehicles},
  author={Zeng, Tengchan and Semiari, Omid and Chen, Mingzhe and Saad, Walid and Bennis, Mehdi},
  journal={arXiv preprint arXiv:2102.03401},
  year={2021}
}

@inproceedings{uchida2017embedding,
  title={Embedding watermarks into deep neural networks},
  author={Uchida, Yusuke and Nagai, Yuki and Sakazawa, Shigeyuki and Satoh, Shin'ichi},
  booktitle={Proceedings of the 2017 ACM on international conference on multimedia retrieval},
  pages={269--277},
  year={2017}
}

@article{oliynyk2023know,
  title={I know what you trained last summer: A survey on stealing machine learning models and defences},
  author={Oliynyk, Daryna and Mayer, Rudolf and Rauber, Andreas},
  journal={ACM Computing Surveys},
  volume={55},
  number={14s},
  pages={1--41},
  year={2023},
  publisher={ACM New York, NY}
}

@inproceedings{adi2018turning,
  title={Turning your weakness into a strength: Watermarking deep neural networks by backdooring},
  author={Adi, Yossi and Baum, Carsten and Cisse, Moustapha and Pinkas, Benny and Keshet, Joseph},
  booktitle={27th USENIX security symposium (USENIX Security 18)},
  pages={1615--1631},
  year={2018}
}

@inproceedings{carlini2022membership,
  title={Membership inference attacks from first principles},
  author={Carlini, Nicholas and Chien, Steve and Nasr, Milad and Song, Shuang and Terzis, Andreas and Tramer, Florian},
  booktitle={2022 IEEE symposium on security and privacy (SP)},
  pages={1897--1914},
  year={2022},
  organization={IEEE}
}

@inproceedings{cheon2017homomorphic,
  title={Homomorphic encryption for arithmetic of approximate numbers},
  author={Cheon, Jung Hee and Kim, Andrey and Kim, Miran and Song, Yongsoo},
  booktitle={International conference on the theory and application of cryptology and information security},
  pages={409--437},
  year={2017},
  organization={Springer}
}

@article{zhu2019deep,
  title={Deep leakage from gradients},
  author={Zhu, Ligeng and Liu, Zhijian and Han, Song},
  journal={Advances in neural information processing systems},
  volume={32},
  year={2019}
}

@misc{qayyum2021collaborative,
      title={Collaborative Federated Learning For Healthcare: Multi-Modal COVID-19 Diagnosis at the Edge}, 
      author={Adnan Qayyum and Kashif Ahmad and Muhammad Ahtazaz Ahsan and Ala Al-Fuqaha and Junaid Qadir},
      year={2021},
      eprint={2101.07511},
      archivePrefix={arXiv},
      primaryClass={cs.LG},
      url={https://arxiv.org/abs/2101.07511}, 
}

@article{dalzochio_machine_2020,
	title = {Machine learning and reasoning for predictive maintenance in Industry 4.0: Current status and challenges},
	volume = {123},
	shorttitle = {Machine learning and reasoning for predictive maintenance in Industry 4.0},
	pages = {103298},
	journal = {Computers in Industry},
	author = {Dalzochio, Jovani and Kunst, Rafael and Pignaton, Edison and Binotto, Alecio and Sanyal, Srijnan and Favilla, Jose and Barbosa, Jorge},
	year = {2020},
	note = {Publisher: Elsevier},
}

@article{liu2018cps,
  title={CPS-based smart warehouse for industry 4.0: a survey of the underlying technologies},
  author={Liu, Xiulong and Cao, Jiannong and Yang, Yanni and Jiang, Shan},
  journal={Computers},
  volume={7},
  number={1},
  pages={13},
  year={2018},
  publisher={Multidisciplinary Digital Publishing Institute},
 }

@article{gupta_smart_2020,
	title = {Smart {Contract} {Privacy} {Protection} {Using} {AI} in {Cyber}-{Physical} {Systems}: {Tools}, {Techniques} and {Challenges}},
	volume = {8},
	issn = {2169-3536},
	shorttitle = {Smart {Contract} {Privacy} {Protection} {Using} {AI} in {Cyber}-{Physical} {Systems}},
	journal = {IEEE Access},
	author = {Gupta, Rajesh and Tanwar, Sudeep and Al-Turjman, Fadi and Italiya, Prit and Nauman, Ali and Kim, Sung Won},
	year = {2020},
	pages = {24746--24772},
}

@article{qian_towards_2018,
	title = {Towards decentralized {IoT} security enhancement: {A} blockchain approach},
	volume = {72},
	issn = {0045-7906},
	shorttitle = {Towards decentralized {IoT} security enhancement},
	language = {en},
	urlyear = {2020-05-11},
	journal = {Computers \& Electrical Engineering},
	author = {Qian, Yongfeng and Jiang, Yingying and Chen, Jing and Zhang, Yu and Song, Jeungeun and Zhou, Ming and Pustišek, Matevž},
	month = nov,
	year = {2018},
	pages = {266--273},
}

@article{younan_challenges_2020,
	title = {Challenges and recommended technologies for the industrial internet of things: {A} comprehensive review},
	volume = {151},
	issn = {0263-2241},
	shorttitle = {Challenges and recommended technologies for the industrial internet of things},
	language = {en},
	urlyear = {2020-05-11},
	journal = {Measurement},
	author = {Younan, Mina and Houssein, Essam H. and Elhoseny, Mohamed and Ali, Abdelmgeid A.},
	month = feb,
	year = {2020},
	pages = {107198},
}

@article{sitton-candanedo_review_2019,
	title = {A review of edge computing reference architectures and a new global edge proposal},
	volume = {99},
	issn = {0167-739X},
	language = {en},
	urlyear = {2020-05-11},
	journal = {Future Generation Computer Systems},
	author = {Sittón-Candanedo, Inés and Alonso, Ricardo S. and Corchado, Juan M. and Rodríguez-González, Sara and Casado-Vara, Roberto},
	month = oct,
	year = {2019},
	pages = {278--294},
}

@article{khan_industrial_2020,
	title = {Industrial internet of things: {Recent} advances, enabling technologies and open challenges},
	volume = {81},
	issn = {0045-7906},
	shorttitle = {Industrial internet of things},
	language = {en},
	urlyear = {2020-05-11},
	journal = {Computers \& Electrical Engineering},
	author = {Khan, W. Z. and Rehman, M. H. and Zangoti, H. M. and Afzal, M. K. and Armi, N. and Salah, K.},
	month = jan,
	year = {2020},
	pages = {106522},
}

@article{penumuru2019identification,
  title={Identification and classification of materials using machine vision and machine learning in the context of industry 4.0},
  author={Penumuru, Durga Prasad and Muthuswamy, Sreekumar and Karumbu, Premkumar},
  journal={Journal of Intelligent Manufacturing},
  volume={31},
  number={5},
  pages={1229--1241},
  year={2020},
  publisher={Springer}
}

@article{myllyaho2022misbehaviour,
  title={On misbehaviour and fault tolerance in machine learning systems},
  author={Myllyaho, Lalli and Raatikainen, Mikko and M{\"a}nnist{\"o}, Tomi and Nurminen, Jukka K and Mikkonen, Tommi},
  journal={Journal of Systems and Software},
  volume={183},
  pages={111096},
  year={2022},
  publisher={Elsevier}
}

@inproceedings{roffe2020evaluation,
  title={Evaluation of algorithm-based fault tolerance for machine learning and computer vision under neutron radiation},
  author={Roffe, Seth and George, Alan D},
  booktitle={2020 IEEE Aerospace Conference},
  pages={1--9},
  year={2020},
  organization={IEEE}
}

@misc{aws,
title = {Amazon Web Services. 2021. Amazon SageMaker},
url = {https://aws.amazon.com/sagemaker/developer-resources/}
}

@misc{googleTPU,
title = {Google. 2021. Google Cloud TPU},
url = {https://cloud.google.com/tpu}
}

@misc{ibmCloud,
title = {\mbox{IBM Cloud}. 2021. \mbox{IBM} \mbox{W}atson \mbox{M}achine \mbox{L}earning},
url = {https://www.ibm.com/cloud/machine-learning}
}

@misc{microsoftAzure,
title = {Microsoft. 2021. Microsoft Azure Machine Learning},
url = {https://azure.microsoft.com/en-us/overview/machine-learning/}
}

@article{kairouz2021advances,
  title={Advances and open problems in federated learning},
  author={Kairouz, Peter and McMahan, H Brendan and Avent, Brendan and Bellet, Aur{\'e}lien and Bennis, Mehdi and Bhagoji, Arjun Nitin and Bonawitz, Kallista and Charles, Zachary and Cormode, Graham and Cummings, Rachel and others},
  journal={Foundations and trends{\textregistered} in machine learning},
  volume={14},
  number={1--2},
  pages={1--210},
  year={2021},
  publisher={Now Publishers, Inc.}
}

@inproceedings{karimireddy2020scaffold,
  title={Scaffold: Stochastic controlled averaging for federated learning},
  author={Karimireddy, Sai Praneeth and Kale, Satyen and Mohri, Mehryar and Reddi, Sashank and Stich, Sebastian and Suresh, Ananda Theertha},
  booktitle={International conference on machine learning},
  pages={5132--5143},
  year={2020},
  organization={PMLR}
}

@article{pillutla2022robust,
  title={Robust aggregation for federated learning},
  author={Pillutla, Krishna and Kakade, Sham M and Harchaoui, Zaid},
  journal={IEEE Transactions on Signal Processing},
  volume={70},
  pages={1142--1154},
  year={2022},
  publisher={IEEE}
}

@article{li2021fedbn,
  title={Fedbn: Federated learning on non-iid features via local batch normalization},
  author={Li, Xiaoxiao and Jiang, Meirui and Zhang, Xiaofei and Kamp, Michael and Dou, Qi},
  journal={arXiv preprint arXiv:2102.07623},
  year={2021}
}

@inproceedings{li2021model,
  title={Model-contrastive federated learning},
  author={Li, Qinbin and He, Bingsheng and Song, Dawn},
  booktitle={Proceedings of the IEEE/CVF conference on computer vision and pattern recognition},
  pages={10713--10722},
  year={2021}
}

@article{jimenez2025thorough,
  title={A thorough assessment of the non-iid data impact in federated learning},
  author={Jimenez-Gutierrez, Daniel M and Hassanzadeh, Mehrdad and Anagnostopoulos, Aris and Chatzigiannakis, Ioannis and Vitaletti, Andrea},
  journal={Journal of Industrial Information Integration},
  pages={101052},
  year={2025},
  publisher={Elsevier}
}

@article{zhao2018federated,
  title={Federated learning with non-iid data},
  author={Zhao, Yue and Li, Meng and Lai, Liangzhen and Suda, Naveen and Civin, Damon and Chandra, Vikas},
  journal={arXiv:1806.00582},
  year={2018}
}

@article{caldas2018leaf,
  title={Leaf: A benchmark for federated settings},
  author={Caldas, Sebastian and Duddu, Sai Meher Karthik and Wu, Peter and Li, Tian and Kone{\v{c}}n{\`y}, Jakub and McMahan, H Brendan and Smith, Virginia and Talwalkar, Ameet},
  journal={arXiv preprint arXiv:1812.01097},
  year={2018}
}

@article{acar2021federated,
  title={Federated learning based on dynamic regularization},
  author={Acar, Durmus Alp Emre and Zhao, Yue and Navarro, Ramon Matas and Mattina, Matthew and Whatmough, Paul N and Saligrama, Venkatesh},
  journal={arXiv preprint arXiv:2111.04263},
  year={2021}
}

@article{reddi2020adaptive,
  title={Adaptive federated optimization},
  author={Reddi, Sashank and Charles, Zachary and Zaheer, Manzil and Garrett, Zachary and Rush, Keith and Kone{\v{c}}n{\`y}, Jakub and Kumar, Sanjiv and McMahan, H Brendan},
  journal={arXiv preprint arXiv:2003.00295},
  year={2020}
}

@article{fallah2020personalized,
  title={Personalized federated learning with theoretical guarantees: A model-agnostic meta-learning approach},
  author={Fallah, Alireza and Mokhtari, Aryan and Ozdaglar, Asuman},
  journal={Advances in neural information processing systems},
  volume={33},
  pages={3557--3568},
  year={2020}
}

@article{t2020personalized,
  title={Personalized federated learning with moreau envelopes},
  author={T Dinh, Canh and Tran, Nguyen and Nguyen, Josh},
  journal={Advances in neural information processing systems},
  volume={33},
  pages={21394--21405},
  year={2020}
}

@inproceedings{li2021ditto,
  title={Ditto: Fair and robust federated learning through personalization},
  author={Li, Tian and Hu, Shengyuan and Beirami, Ahmad and Smith, Virginia},
  booktitle={International conference on machine learning},
  pages={6357--6368},
  year={2021},
  organization={PMLR}
}

@article{chen2025effective,
  title={Effective Non-IID Degree Estimation for Robust Federated Learning in Healthcare Datasets},
  author={Chen, Kun-Yi and Shyu, Chi-Ren and Tsai, Yuan-Yu and Baskett, William I and Chang, Chi-Yu and Chou, Che-Yi and Tsai, Jeffrey JP and Shae, Zon-Yin},
  journal={Journal of Healthcare Informatics Research},
  pages={1--28},
  year={2025},
  publisher={Springer}
}

@inproceedings{abadi2016deep,
  title={Deep learning with differential privacy},
  author={Abadi, Martin and Chu, Andy and Goodfellow, Ian and McMahan, H Brendan and Mironov, Ilya and Talwar, Kunal and Zhang, Li},
  booktitle={Proceedings of the 2016 ACM SIGSAC conference on computer and communications security},
  pages={308--318},
  year={2016}
}

@article{papernot2018scalable,
  title={Scalable private learning with pate},
  author={Papernot, Nicolas and Song, Shuang and Mironov, Ilya and Raghunathan, Ananth and Talwar, Kunal and Erlingsson, {\'U}lfar},
  journal={arXiv preprint arXiv:1802.08908},
  year={2018}
}

@article{bernstein2018signsgd,
  title={signSGD with majority vote is communication efficient and fault tolerant},
  author={Bernstein, Jeremy and Zhao, Jiawei and Azizzadenesheli, Kamyar and Anandkumar, Anima},
  journal={arXiv preprint arXiv:1810.05291},
  year={2018}
}

@inproceedings{bonawitz2017practical,
  title={Practical secure aggregation for privacy-preserving machine learning},
  author={Bonawitz, Keith and Ivanov, Vladimir and Kreuter, Ben and Marcedone, Antonio and McMahan, H Brendan and Patel, Sarvar and Ramage, Daniel and Segal, Aaron and Seth, Karn},
  booktitle={proceedings of the 2017 ACM SIGSAC Conference on Computer and Communications Security},
  pages={1175--1191},
  year={2017}
}

@inproceedings{mohassel2018aby3,
  title={ABY3: A mixed protocol framework for machine learning},
  author={Mohassel, Payman and Rindal, Peter},
  booktitle={Proceedings of the 2018 ACM SIGSAC conference on computer and communications security},
  pages={35--52},
  year={2018}
}

@article{knott2021crypten,
  title={Crypten: Secure multi-party computation meets machine learning},
  author={Knott, Brian and Venkataraman, Shobha and Hannun, Awni and Sengupta, Shubho and Ibrahim, Mark and van der Maaten, Laurens},
  journal={Advances in Neural Information Processing Systems},
  volume={34},
  pages={4961--4973},
  year={2021}
}

@inproceedings{ohrimenko2016oblivious,
  title={Oblivious $\{$Multi-Party$\}$ machine learning on trusted processors},
  author={Ohrimenko, Olga and Schuster, Felix and Fournet, C{\'e}dric and Mehta, Aastha and Nowozin, Sebastian and Vaswani, Kapil and Costa, Manuel},
  booktitle={25th USENIX Security Symposium (USENIX Security 16)},
  pages={619--636},
  year={2016}
}

@Article{wu2025survey,
AUTHOR = {Wu, Lingling and others},
TITLE = {Homomorphic Encryption for Machine Learning Applications with CKKS Algorithms: A Survey of Developments and Applications},
JOURNAL = {Computers, Materials \& Continua},
VOLUME = {85},
YEAR = {2025},
NUMBER = {1},
PAGES = {89--119},
URL = {http://www.techscience.com/cmc/v85n1/63510},
ISSN = {1546-2226},
DOI = {10.32604/cmc.2025.064346}
}

@inproceedings{fang2020local,
  title={Local model poisoning attacks to $\{$Byzantine-Robust$\}$ federated learning},
  author={Fang, Minghong and Cao, Xiaoyu and Jia, Jinyuan and Gong, Neil},
  booktitle={29th USENIX security symposium (USENIX Security 20)},
  pages={1605--1622},
  year={2020}
}

@inproceedings{bagdasaryan2020backdoor,
  title={How to backdoor federated learning},
  author={Bagdasaryan, Eugene and Veit, Andreas and Hua, Yiqing and Estrin, Deborah and Shmatikov, Vitaly},
  booktitle={International conference on artificial intelligence and statistics},
  pages={2938--2948},
  year={2020},
  organization={PMLR}
}

@inproceedings{bruckner2011stackelberg,
  title={Stackelberg games for adversarial prediction problems},
  author={Br{\"u}ckner, Michael and Scheffer, Tobias},
  booktitle={Proceedings of the 17th ACM SIGKDD international conference on Knowledge discovery and data mining},
  pages={547--555},
  year={2011}
}

@article{chen2020learning,
  title={Learning strategy-aware linear classifiers},
  author={Chen, Yiling and Liu, Yang and Podimata, Chara},
  journal={Advances in Neural Information Processing Systems},
  volume={33},
  pages={15265--15276},
  year={2020}
}

@inproceedings{liu2024game,
  title={Game-theoretic unlearnable example generator},
  author={Liu, Shuang and Wang, Yihan and Gao, Xiao-Shan},
  booktitle={Proceedings of the AAAI Conference on Artificial Intelligence},
  volume={38},
  number={19},
  pages={21349--21358},
  year={2024}
}

@article{wang2024rfvir,
  title={RFVIR: A robust federated algorithm defending against Byzantine attacks},
  author={Wang, Yongkang and Zhai, Di-Hua and Xia, Yuanqing},
  journal={Information Fusion},
  volume={105},
  pages={102251},
  year={2024},
  publisher={Elsevier}
}

@article{nguyen2024backdoor,
  title={Backdoor attacks and defenses in federated learning: Survey, challenges and future research directions},
  author={Nguyen, Thuy Dung and Nguyen, Tuan and Le Nguyen, Phi and Pham, Hieu H and Doan, Khoa D and Wong, Kok-Seng},
  journal={Engineering Applications of Artificial Intelligence},
  volume={127},
  pages={107166},
  year={2024},
  publisher={Elsevier}
}

@article{chai2024survey,
  title={A survey for federated learning evaluations: Goals and measures},
  author={Chai, Di and Wang, Leye and Yang, Liu and Zhang, Junxue and Chen, Kai and Yang, Qiang},
  journal={IEEE transactions on knowledge and data engineering},
  volume={36},
  number={10},
  pages={5007--5024},
  year={2024},
  publisher={IEEE}
}

@inproceedings{carletti2025sok,
  title={$\{$SoK$\}$: Gradient Inversion Attacks in Federated Learning},
  author={Carletti, Vincenzo and Foggia, Pasquale and Mazzocca, Carlo and Parrella, Giuseppe and Vento, Mario},
  booktitle={34th USENIX Security Symposium (USENIX Security 25)},
  pages={6439--6459},
  year={2025}
}

@article{leite2024federated,
  title={Federated learning under attack: Improving gradient inversion for batch of images},
  author={Leite, Luiz and Santo, Yuri and Dalmazo, Bruno L and Riker, Andr{\'e}},
  journal={arXiv preprint arXiv:2409.17767},
  year={2024}
}

@article{yuan2025approximate,
  title={Approximate homomorphic encryption based privacy-preserving machine learning: a survey},
  author={Yuan, Jiangjun and Liu, Weinan and Shi, Jiawen and Li, Qingqing},
  journal={Artificial Intelligence Review},
  volume={58},
  number={3},
  pages={82},
  year={2025},
  publisher={Springer}
}

@article{tan2024long,
  author={Tan, Jinghong and Liu, Zhian and Guo, Kun and Zhao, Mingxiong},
  journal={IEEE Internet of Things Journal}, 
  title={Long-Term Client Selection for Federated Learning With Non-IID Data: A Truthful Auction Approach}, 
  year={2025},
  volume={12},
  number={5},
  pages={4953-4970},
  doi={10.1109/JIOT.2024.3524389}}

@article{yousefpour2023green,
  title={Green federated learning},
  author={Yousefpour, Ashkan and Guo, Shen and Shenoy, Ashish and Ghosh, Sayan and Stock, Pierre and Maeng, Kiwan and Kr{\"u}ger, Schalk-Willem and Rabbat, Michael and Wu, Carole-Jean and Mironov, Ilya},
  journal={arXiv preprint arXiv:2303.14604},
  year={2023}
}

@article{thakur2025green,
  title={Green federated learning: A new era of green aware AI},
  author={Thakur, Dipanwita and Guzzo, Antonella and Fortino, Giancarlo and Piccialli, Francesco},
  journal={ACM Computing Surveys},
  volume={57},
  number={8},
  pages={1--36},
  year={2025},
  publisher={ACM New York, NY}
}

@inproceedings{smuha2025regulation,
  author       = {Abbasi, A.},
  title        = {Carbon-aware Federated Learning with Model Size Adaptation},
  booktitle    = {Proceedings of the 2024 IEEE International Conference on},
  year         = {2024},
}

@article{ai2023artificial,
  author      = {Tabassi, Elham},
  title       = {Artificial Intelligence Risk Management Framework ({AI RMF} 1.0)},
  institution = {National Institute of Standards and Technology},
  address     = {Gaithersburg, MD},
  type        = {NIST Trustworthy and Responsible AI},
  number      = {NIST AI 100-1},
  year        = {2023},
  doi         = {10.6028/NIST.AI.100-1},
  url         = {https://tsapps.nist.gov/publication/get_pdf.cfm?pub_id=936225},
  note        = {Accessed: October 5, 2026}
}

@article{shoeybi2019megatron,
  title={Megatron-lm: Training multi-billion parameter language models using model parallelism},
  author={Shoeybi, Mohammad and Patwary, Mostofa and Puri, Raul and LeGresley, Patrick and Casper, Jared and Catanzaro, Bryan},
  journal={arXiv preprint arXiv:1909.08053},
  year={2019}
}

@article{huang2019gpipe,
  title={Gpipe: Efficient training of giant neural networks using pipeline parallelism},
  author={Huang, Yanping and Cheng, Youlong and Bapna, Ankur and Firat, Orhan and Chen, Dehao and Chen, Mia and Lee, HyoukJoong and Ngiam, Jiquan and Le, Quoc V and Wu, Yonghui and others},
  journal={Advances in neural information processing systems},
  volume={32},
  year={2019}
}

@inproceedings{narayanan2021efficient,
  title={Efficient large-scale language model training on gpu clusters using megatron-lm},
  author={Narayanan, Deepak and Shoeybi, Mohammad and Casper, Jared and LeGresley, Patrick and Patwary, Mostofa and Korthikanti, Vijay and Vainbrand, Dmitri and Kashinkunti, Prethvi and Bernauer, Julie and Catanzaro, Bryan and others},
  booktitle={Proceedings of the international conference for high performance computing, networking, storage and analysis},
  pages={1--15},
  year={2021}
}

@inproceedings{liu2025data,
  title={Data Heterogeneity Modeling for Trustworthy Machine Learning},
  author={Liu, Jiashuo and Cui, Peng},
  booktitle={Proceedings of the 31st ACM SIGKDD Conference on Knowledge Discovery and Data Mining V. 2},
  pages={6086--6095},
  year={2025}
}

@article{chen2016revisiting,
  title={Revisiting distributed synchronous SGD},
  author={Chen, Jianmin and Pan, Xinghao and Monga, Rajat and Bengio, Samy and Jozefowicz, Rafal},
  journal={arXiv preprint arXiv:1604.00981},
  year={2016}
}

@inproceedings{sapio2021scaling,
  title={Scaling distributed machine learning with $\{$In-Network$\}$ aggregation},
  author={Sapio, Amedeo and Canini, Marco and Ho, Chen-Yu and Nelson, Jacob and Kalnis, Panos and Kim, Changhoon and Krishnamurthy, Arvind and Moshref, Masoud and Ports, Dan and Richt{\'a}rik, Peter},
  booktitle={18th USENIX Symposium on Networked Systems Design and Implementation (NSDI 21)},
  pages={785--808},
  year={2021}
}

@inproceedings{cui2014exploiting,
  title={Exploiting bounded staleness to speed up big data analytics},
  author={Cui, Henggang and Cipar, James and Ho, Qirong and Kim, Jin Kyu and Lee, Seunghak and Kumar, Abhimanu and Wei, Jinliang and Dai, Wei and Ganger, Gregory R and Gibbons, Phillip B and others},
  booktitle={2014 USENIX Annual Technical Conference (USENIX ATC 14)},
  pages={37--48},
  year={2014}
}

@article{teo2024federated,
  title={Federated machine learning in healthcare: A systematic review on clinical applications and technical architecture},
  author={Teo, Zhen Ling and Jin, Liyuan and Liu, Nan and Li, Siqi and Miao, Di and Zhang, Xiaoman and Ng, Wei Yan and Tan, Ting Fang and Lee, Deborah Meixuan and Chua, Kai Jie and others},
  journal={Cell Reports Medicine},
  volume={5},
  number={2},
  year={2024},
  publisher={Elsevier}
}

@inproceedings{tramer2016stealing,
  title={Stealing machine learning models via prediction $\{$APIs$\}$},
  author={Tram{\`e}r, Florian and Zhang, Fan and Juels, Ari and Reiter, Michael K and Ristenpart, Thomas},
  booktitle={25th USENIX security symposium (USENIX Security 16)},
  pages={601--618},
  year={2016}
}

@inproceedings{shokri2017membership,
  title={Membership inference attacks against machine learning models},
  author={Shokri, Reza and Stronati, Marco and Song, Congzheng and Shmatikov, Vitaly},
  booktitle={2017 IEEE symposium on security and privacy (SP)},
  pages={3--18},
  year={2017},
  organization={IEEE}
}

@article{geiping2020inverting,
  title={Inverting gradients-how easy is it to break privacy in federated learning?},
  author={Geiping, Jonas and Bauermeister, Hartmut and Dr{\"o}ge, Hannah and Moeller, Michael},
  journal={Advances in neural information processing systems},
  volume={33},
  pages={16937--16947},
  year={2020}
}

@article{fung2018mitigating,
  title={Mitigating sybils in federated learning poisoning},
  author={Fung, Clement and Yoon, Chris JM and Beschastnikh, Ivan},
  journal={arXiv preprint arXiv:1808.04866},
  year={2018}
}

@inproceedings{yin2018byzantine,
  title={Byzantine-robust distributed learning: Towards optimal statistical rates},
  author={Yin, Dong and Chen, Yudong and Kannan, Ramchandran and Bartlett, Peter},
  booktitle={International conference on machine learning},
  pages={5650--5659},
  year={2018},
  organization={Pmlr}
}

@article{blanchard2017machine,
  title={Machine learning with adversaries: Byzantine tolerant gradient descent},
  author={Blanchard, Peva and El Mhamdi, El Mahdi and Guerraoui, Rachid and Stainer, Julien},
  journal={Advances in neural information processing systems},
  volume={30},
  year={2017}
}

@article{uddin2025systematic,
  title={A systematic literature review of robust federated learning: Issues, solutions, and future research directions},
  author={Uddin, Md Palash and Xiang, Yong and Hasan, Mahmudul and Bai, Jun and Zhao, Yao and Gao, Longxiang},
  journal={ACM Computing Surveys},
  volume={57},
  number={10},
  pages={1--62},
  year={2025},
  publisher={ACM New York, NY}
}

@article{wan2024data,
  title={Data and model poisoning backdoor attacks on wireless federated learning, and the defense mechanisms: A comprehensive survey},
  author={Wan, Yichen and Qu, Youyang and Ni, Wei and Xiang, Yong and Gao, Longxiang and Hossain, Ekram},
  journal={IEEE Communications Surveys \& Tutorials},
  volume={26},
  number={3},
  pages={1861--1897},
  year={2024},
  publisher={IEEE}
}

@inproceedings{li2022federated,
  title={Federated learning on non-iid data silos: An experimental study},
  author={Li, Qinbin and Diao, Yiqun and Chen, Quan and He, Bingsheng},
  booktitle={2022 IEEE 38th international conference on data engineering (ICDE)},
  pages={965--978},
  year={2022},
  organization={IEEE}
}

@article{nikolic2026model,
 author = {Nikolic, Ivica and Baluta, Teodora and Saxena, Prateek},
 booktitle = {Advances in Neural Information Processing Systems},
 doi = {10.52202/085713-1146},
 editor = {D. Belgrave and C. Zhang and H. Lin and R. Pascanu and P. Koniusz and M. Ghassemi and N. Chen},
 pages = {34126--34153},
 publisher = {Curran Associates, Inc.},
 title = {Model Provenance Testing for Large Language Models},
 url = {https://proceedings.neurips.cc/paper_files/paper/2025/file/30e55dfbfa8e76e3526de7f2ab029cda-Paper-Conference.pdf},
 volume = {38, Main Conference},
 year = {2025}
}

@article{li2023experimental,
  author={Li, Shenghui and Ngai, Edith C.-H. and Voigt, Thiemo},
  journal={IEEE Transactions on Big Data}, 
  title={An Experimental Study of Byzantine-Robust Aggregation Schemes in Federated Learning}, 
  year={2024},
  volume={10},
  number={6},
  pages={975-988},
  doi={10.1109/TBDATA.2023.3237397}}

@article{chen2023privacy,
  title={Privacy and fairness in federated learning: On the perspective of tradeoff},
  author={Chen, Huiqiang and Zhu, Tianqing and Zhang, Tao and Zhou, Wanlei and Yu, Philip S},
  journal={ACM Computing Surveys},
  volume={56},
  number={2},
  pages={1--37},
  year={2023},
  publisher={ACM New York, NY}
}

@article{tanveer2025balancing,
  title={Balancing privacy and performance in healthcare: A federated learning framework for sensitive data},
  author={Tanveer, Fatima and Iradat, Faisal and Iqbal, Waseem and Alsagri, Hatoon S and Alhakbani, Haya Abdullah A and Ahmad, Awais and Khan, Fakhri Alam},
  journal={Digital Health},
  volume={11},
  pages={20552076251381769},
  year={2025},
  publisher={SAGE Publications Sage UK: London, England}
}

@inproceedings{tekgul2021waffle,
  title={Waffle: Watermarking in federated learning},
  author={Tekgul, Buse GA and Xia, Yuxi and Marchal, Samuel and Asokan, N},
  booktitle={2021 40th International Symposium on Reliable Distributed Systems (SRDS)},
  pages={310--320},
  year={2021},
  organization={IEEE}
}

@article{yang2023PrivacyandModelIP,
  title={Federated Learning with Privacy-preserving and Model IP-right-protection},
  author={Yang, Qiang and Huang, Anbu and Fan, Lixin and Chan, Chee Seng and Lim, Jian Han and Ng, Kam Woh and Ong, Ding Sheng and Li, Bowen},
  journal={Machine Intelligence Research},
  volume={20},
  number={1},
  pages={19--37},
  year={2023},
  publisher={Springer}
}

@article{sanchez2024federatedtrust,
  title={Federatedtrust: A solution for trustworthy federated learning},
  author={S{\'a}nchez, Pedro Miguel S{\'a}nchez and Celdr{\'a}n, Alberto Huertas and Xie, Ning and Bovet, G{\'e}r{\^o}me and P{\'e}rez, Gregorio Mart{\'\i}nez and Stiller, Burkhard},
  journal={Future Generation Computer Systems},
  volume={152},
  pages={83--98},
  year={2024},
  publisher={Elsevier}
}

@article{wagner2018technical,
  title={Technical privacy metrics: a systematic survey},
  author={Wagner, Isabel and Eckhoff, David},
  journal={ACM Computing Surveys (Csur)},
  volume={51},
  number={3},
  pages={1--38},
  year={2018},
  publisher={ACM New York, NY, USA}
}

@article{zhang2023trading,
  title={Trading off privacy, utility, and efficiency in federated learning},
  author={Zhang, Xiaojin and Kang, Yan and Chen, Kai and Fan, Lixin and Yang, Qiang},
  journal={ACM Transactions on Intelligent Systems and Technology},
  volume={14},
  number={6},
  pages={1--32},
  year={2023},
  publisher={ACM New York, NY}
}

@article{kim2024spafl,
  title={SpaFL: Communication-efficient federated learning with sparse models and low computational overhead},
  author={Kim, Minsu and Saad, Walid and Debbah, Merouane and Hong, Choong S},
  journal={Advances in Neural Information Processing Systems},
  volume={37},
  pages={86500--86527},
  year={2024}
}

@inproceedings{xia2024towards,
  title={Towards energy-aware federated learning via marl: A dual-selection approach for model and client},
  author={Xia, Jun and Zhang, Yi and Shi, Yiyu},
  booktitle={Proceedings of the 43rd IEEE/ACM International Conference on Computer-Aided Design},
  pages={1--9},
  year={2024}
}

@article{koloskova2022sharper,
  title={Sharper convergence guarantees for asynchronous SGD for distributed and federated learning},
  author={Koloskova, Anastasiia and Stich, Sebastian U and Jaggi, Martin},
  journal={Advances in Neural Information Processing Systems},
  volume={35},
  pages={17202--17215},
  year={2022}
}

@inproceedings{thapa2022splitfed,
  title={Splitfed: When federated learning meets split learning},
  author={Thapa, Chandra and Arachchige, Pathum Chamikara Mahawaga and Camtepe, Seyit and Sun, Lichao},
  booktitle={Proceedings of the AAAI conference on artificial intelligence},
  volume={36},
  number={8},
  pages={8485--8493},
  year={2022}
}

@article{khan2021federated,
  title={Federated learning for internet of things: Recent advances, taxonomy, and open challenges},
  author={Khan, Latif U and Saad, Walid and Han, Zhu and Hossain, Ekram and Hong, Choong Seon},
  journal={IEEE Communications Surveys \& Tutorials},
  volume={23},
  number={3},
  pages={1759--1799},
  year={2021},
  publisher={IEEE}
}

@article{li2024holistic,
  title={Holistic evaluation metrics: Use case sensitive evaluation metrics for federated learning},
  author={Li, Yanli and Ibrahim, Jehad and Chen, Huaming and Yuan, Dong and Choo, Kim-Kwang Raymond},
  journal={arXiv:2405.02360},
  year={2024}
}

@article{thein2024personalized,
  title={Personalized federated learning-based intrusion detection system: Poisoning attack and defense},
  author={Thein, Thin Tharaphe and Shiraishi, Yoshiaki and Morii, Masakatu},
  journal={Future Generation Computer Systems},
  volume={153},
  pages={182--192},
  year={2024},
  publisher={Elsevier}
}

@article{zeng2024bsr,
  title={Bsr-fl: An efficient byzantine-robust privacy-preserving federated learning framework},
  author={Zeng, Honghong and Li, Jie and Lou, Jiong and Yuan, Shijing and Wu, Chentao and Zhao, Wei and Wu, Sijin and Wang, Zhiwen},
  journal={IEEE Transactions on Computers},
  volume={73},
  number={8},
  pages={2096--2110},
  year={2024},
  publisher={IEEE}
}

@inproceedings{zawad2021curse,
  title={Curse or redemption? how data heterogeneity affects the robustness of federated learning},
  author={Zawad, Syed and Ali, Ahsan and Chen, Pin-Yu and Anwar, Ali and Zhou, Yi and Baracaldo, Nathalie and Tian, Yuan and Yan, Feng},
  booktitle={Proceedings of the AAAI conference on artificial intelligence},
  volume={35},
  number={12},
  pages={10807--10814},
  year={2021}
}

@article{Gutierrez2024NonIID,
  title={Non-IID data in Federated Learning: A Survey with Taxonomy, Metrics, Methods, Frameworks and Future Directions},
  author={Daniel Mauricio Jimenez Gutierrez and David Solans and Mikko A. Heikkil{\"a} and Andrea Vitaletti and Nicolas Kourtellis and Aris Anagnostopoulos and Ioannis Chatzigiannakis},
  journal={arXiv:2411.12377},
  year={2024},
}

@inproceedings{le2025fedklentropy,
  title={FedKLEntropy: Enhancing Federated Learning via Model Entropy and Kullback-Leibler Divergence},
  author={Le, Duy-Dong and Huynh, Tuong-Nguyen and Le, Nhat Tung},
  booktitle={Proceedings of the International Conference on Research in Adaptive and Convergent Systems},
  pages={1--8},
  year={2025}
}

@inproceedings{alharbi2023robust,
  title={Robust Federated Learning Method Against Data and Model Poisoning Attacks with Heterogeneous Data Distribution.},
  author={Alharbi, Ebtisaam and Marcolino, Leandro Soriano and Gouglidis, Antonios and Ni, Qiang},
  booktitle={ECAI},
  pages={85--92},
  year={2023}
}

@article{he2020fedml,
  title={Fedml: A research library and benchmark for federated machine learning},
  author={He, Chaoyang and Li, Songze and So, Jinhyun and Zeng, Xiao and Zhang, Mi and Wang, Hongyi and Wang, Xiaoyang and Vepakomma, Praneeth and Singh, Abhishek and Qiu, Hang and others},
  journal={arXiv:2007.13518},
  year={2020}
}

@article{lai2022fedscale,
  author       = {Goussakov, R.},
  title        = {Hellinger Distance-based Similarity Measures for Recommender Systems},
  year         = {2020},
}

@inproceedings{feng2023fedmultimodal,
  title={Fedmultimodal: A benchmark for multimodal federated learning},
  author={Feng, Tiantian and Bose, Digbalay and Zhang, Tuo and Hebbar, Rajat and Ramakrishna, Anil and Gupta, Rahul and Zhang, Mi and Avestimehr, Salman and Narayanan, Shrikanth},
  booktitle={Proceedings of the 29th ACM SIGKDD conference on knowledge discovery and data mining},
  pages={4035--4045},
  year={2023}
}

@inproceedings{han2024fedsecurity,
  title={Fedsecurity: A benchmark for attacks and defenses in federated learning and federated llms},
  author={Han, Shanshan and Buyukates, Baturalp and Hu, Zijian and Jin, Han and Jin, Weizhao and Sun, Lichao and Wang, Xiaoyang and Wu, Wenxuan and Xie, Chulin and Yao, Yuhang and others},
  booktitle={Proceedings of the 30th ACM SIGKDD Conference on Knowledge Discovery and Data Mining},
  pages={5070--5081},
  year={2024}
}

@article{zhang2025flpoison,
  title={Sok: Benchmarking poisoning attacks and defenses in federated learning},
  author={Zhang, Heyi and Liu, Yule and He, Xinlei and Wu, Jun and Cong, Tianshuo and Huang, Xinyi},
  journal={arXiv:2502.03801},
  year={2025}
}

@inproceedings{zhang2024flhetbench,
  title={Flhetbench: Benchmarking device and state heterogeneity in federated learning},
  author={Zhang, Junyuan and Zeng, Shuang and Zhang, Miao and Wang, Runxi and Wang, Feifei and Zhou, Yuyin and Liang, Paul Pu and Qu, Liangqiong},
  booktitle={Proceedings of the IEEE/CVF Conference on Computer Vision and Pattern Recognition},
  pages={12098--12108},
  year={2024}
}

@article{ogier2022flamby,
  title={Flamby: Datasets and benchmarks for cross-silo federated learning in realistic healthcare settings},
  author={Ogier du Terrail, Jean and Ayed, Samy-Safwan and Cyffers, Edwige and Grimberg, Felix and He, Chaoyang and Loeb, Regis and Mangold, Paul and Marchand, Tanguy and Marfoq, Othmane and Mushtaq, Erum and others},
  journal={Advances in Neural Information Processing Systems},
  volume={35},
  pages={5315--5334},
  year={2022}
}

@article{hu2022oarf,
  title={The oarf benchmark suite: Characterization and implications for federated learning systems},
  author={Hu, Sixu and Li, Yuan and Liu, Xu and Li, Qinbin and Wu, Zhaomin and He, Bingsheng},
  journal={ACM Transactions on Intelligent Systems and Technology (TIST)},
  volume={13},
  number={4},
  pages={1--32},
  year={2022},
  publisher={ACM New York, NY}
}

@article{rodriguez2023survey,
  title={Survey on federated learning threats: Concepts, taxonomy on attacks and defences, experimental study and challenges},
  author={Rodr{\'\i}guez-Barroso, Nuria and Jim{\'e}nez-L{\'o}pez, Daniel and Luz{\'o}n, M Victoria and Herrera, Francisco and Mart{\'\i}nez-C{\'a}mara, Eugenio},
  journal={Information Fusion},
  volume={90},
  pages={148--173},
  year={2023},
  publisher={Elsevier}
}

@article{zhang2024survey,
  title={A survey of trustworthy federated learning: Issues, solutions, and challenges},
  author={Zhang, Yifei and Zeng, Dun and Luo, Jinglong and Fu, Xinyu and Chen, Guanzhong and Xu, Zenglin and King, Irwin},
  journal={ACM Transactions on Intelligent Systems and Technology},
  volume={15},
  number={6},
  pages={1--47},
  year={2024},
  publisher={ACM New York, NY, USA}
}

@article{feng2025survey,
  title={A survey of security threats in federated learning},
  author={Feng, Yunhao and Guo, Yanming and Hou, Yinjian and Wu, Yulun and Lao, Mingrui and Yu, Tianyuan and Liu, Gang},
  journal={Complex \& Intelligent Systems},
  volume={11},
  number={2},
  pages={165},
  year={2025},
  publisher={Springer}
}

@article{10.1145/3724113,
author = {Zhao, Joshua and Bagchi, Saurabh and Avestimehr, Salman and Chan, Kevin and Chaterji, Somali and Dimitriadis, Dimitris and Li, Jiacheng and Li, Ninghui and Nourian, Arash and Roth, Holger},
title = {The Federation Strikes Back: A Survey of Federated Learning Privacy Attacks, Defenses, Applications, and Policy Landscape},
year = {2025},
issue_date = {September 2025},
publisher = {Association for Computing Machinery},
address = {New York, NY, USA},
volume = {57},
number = {9},
issn = {0360-0300},
url = {https://doi.org/10.1145/3724113},
doi = {10.1145/3724113},
journal = {ACM Comput. Surv.},
month = apr,
articleno = {230},
numpages = {37}
}

@article{li2025backdoor,
  title={Backdoor attacks and defense mechanisms in federated learning: A survey},
  author={Li, Zhaozheng and Lan, Jiahe and Yan, Zheng and Gelenbe, Erol},
  journal={Information Fusion},
  volume={123},
  pages={103248},
  year={2025},
  publisher={Elsevier}
}

@article{zhang2025review,
  title={A review of research on secure aggregation for federated learning},
  author={Zhang, Xing and Luo, Yuexiang and Li, Tianning},
  journal={Future Internet},
  volume={17},
  number={7},
  pages={308},
  year={2025},
  publisher={MDPI}
}

@article{mahmud2026survey,
  title={A Survey of Federated Learning: Advances in Architecture, Synchronization, and Security Threats.},
  author={Mahmud, Faisal and Mahmud, Fahim and Rahman, Rashedur M},
  journal={Computers, Materials \& Continua},
  volume={86},
  number={3},
  year={2026}
}

@inproceedings{fraboni2021free,
  title={Free-rider attacks on model aggregation in federated learning},
  author={Fraboni, Yann and Vidal, Richard and Lorenzi, Marco},
  booktitle={International conference on artificial intelligence and statistics},
  pages={1846--1854},
  year={2021},
  organization={PMLR}
}

@article{patel2025towards,
author = {Patel, Raj and Tripathi, Himanshu and Stone, Jasper and Amiri Golilarz, Noorbakhsh and Mittal, Sudip and Rahimi, Shahram and Chaudhary, Vini},
title = {Towards Secure MLOps: Surveying Attacks, Mitigation Strategies, and Research Challenges},
year = {2026},
publisher = {Association for Computing Machinery},
address = {New York, NY, USA},
url = {https://doi.org/10.1145/3796728},
doi = {10.1145/3796728},
note = {Just Accepted},
journal = {ACM Trans. AI Secur. Priv.},
month = mar
}

@article{carlini2019evaluating,
  title={On evaluating adversarial robustness},
  author={Carlini, Nicholas and Athalye, Anish and Papernot, Nicolas and Brendel, Wieland and Rauber, Jonas and Tsipras, Dimitris and Goodfellow, Ian and Madry, Aleksander and Kurakin, Alexey},
  journal={arXiv preprint arXiv:1902.06705},
  year={2019}
}

@article{chen2024trustworthy,
  title={Trustworthy federated learning: privacy, security, and beyond},
  author={Chen, Chunlu and Liu, Ji and Tan, Haowen and Li, Xingjian and Wang, Kevin I-Kai and Li, Peng and Sakurai, Kouichi and Dou, Dejing},
  journal={arXiv preprint arXiv:2411.01583},
  year={2024}
}

@article{kyriakidou2026toward,
  title={Toward a Federated Organizational Intelligence Capability Model: Cross-Silo Federated Learning as a Distributed Dynamic Capability},
  author={Kyriakidou-Zacharoudiou, Avgousta and Tsappi, Elena and Georgiades, Michael},
  journal={Sustainability},
  volume={18},
  number={10},
  pages={4762},
  year={2026},
  publisher={MDPI}
}

@article{pfeiffer2023federated,
  title={Federated learning for computationally constrained heterogeneous devices: A survey},
  author={Pfeiffer, Kilian and Rapp, Martin and Khalili, Ramin and Henkel, J{\"o}rg},
  journal={ACM Computing Surveys},
  volume={55},
  number={14s},
  pages={1--27},
  year={2023},
  publisher={ACM New York, NY}
}

@article{korneev2025survey,
  title   = {A Survey on Verifiable Cross-Silo Federated Learning},
  author  = {Korneev, Aleksei and Ramon, Jan},
  journal = {Transactions on Machine Learning Research},
  year    = {2025},
  month   = jun,
  url     = {https://openreview.net/forum?id=uMir8UIHST}
}

@article{quan2025federated,
  title={Federated learning for cyber physical systems: a comprehensive survey},
  author={Quan, Minh K and Pathirana, Pubudu N and Wijayasundara, Mayuri and Setunge, Sujeeva and Nguyen, Dinh C and Brinton, Christopher G and Love, David J and Poor, H Vincent},
  journal={IEEE Communications Surveys \& Tutorials},
  year={2025},
  publisher={IEEE}
}

@article{goudarzi2021distributed,
  author={Goudarzi, Mohammad and Palaniswami, Marimuthu and Buyya, Rajkumar},
  journal={IEEE Transactions on Mobile Computing}, 
  title={A Distributed Deep Reinforcement Learning Technique for Application Placement in Edge and Fog Computing Environments}, 
  year={2023},
  volume={22},
  number={5},
  pages={2491-2505},
  doi={10.1109/TMC.2021.3123165}}

@article{al2024computing,
  title={The computing continuum: From IoT to the cloud},
  author={Al-Dulaimy, Auday and Jansen, Matthijs and Johansson, Bjarne and Trivedi, Animesh and Iosup, Alexandru and Ashjaei, Mohammad and Galletta, Antonino and Kimovski, Dragi and Prodan, Radu and Tserpes, Konstantinos and others},
  journal={Internet of Things},
  volume={27},
  pages={101272},
  year={2024},
  publisher={Elsevier}
}

@article{mahadevappa2024analyzing,
  title={Analyzing threats and attacks in edge data analytics within IoT environments},
  author={Mahadevappa, Poornima and Al-amri, Redhwan and Alkawsi, Gamal and Alkahtani, Ammar Ahmed and Alghenaim, Mohammed Fahad and Alsamman, Mohammed},
  journal={IoT},
  volume={5},
  number={1},
  pages={123--154},
  year={2024},
  publisher={MDPI}
}

@article{li2021byzantine,
  title={Byzantine-robust aggregation in federated learning empowered industrial IoT},
  author={Li, Shenghui and Ngai, Edith and Voigt, Thiemo},
  journal={IEEE Transactions on Industrial Informatics},
  volume={19},
  number={2},
  pages={1165--1175},
  year={2023},
  publisher={IEEE}
}

@article{angueira2022survey,
  title={A survey of physical layer techniques for secure wireless communications in industry},
  author={Angueira, Pablo and Val, I{\~n}aki and Montalban, Jon and Seijo, {\'O}scar and Iradier, Eneko and Fontaneda, Pablo Sanz and Fanari, Lorenzo and Arriola, Aitor},
  journal={IEEE Communications Surveys \& Tutorials},
  volume={24},
  number={2},
  pages={810--838},
  year={2022},
  publisher={IEEE}
}

@article{alhawas2026federated,
  title={Federated learning in edge computing: Vulnerabilities, attacks, and defenses—a survey},
  author={Alhawas, Sahar and Rassam, Murad A},
  journal={Sensors (Basel, Switzerland)},
  volume={26},
  number={4},
  pages={1275},
  year={2026}
}

@article{butun2019security,
  author={Butun, Ismail and Österberg, Patrik and Song, Houbing},
  journal={IEEE Communications Surveys \& Tutorials}, 
  title={Security of the Internet of Things: Vulnerabilities, Attacks, and Countermeasures}, 
  year={2020},
  volume={22},
  number={1},
  pages={616-644},
  doi={10.1109/COMST.2019.2953364}}

@article{mendes2026survey,
  title={Survey on Federated Learning in RIS, STAR-RIS, and BD-RIS-Assisted Wireless Networks},
  author={Mendes, Isabel Francine and Duarte, Diana Laura F and Guedes, Luiz Gustavo B and Souza, Lucas S and Silva, Rafael S and Souto, Victoria DP and Souza, Richard D},
  journal={IEEE Open Journal of the Communications Society},
  year={2026},
  publisher={IEEE}
}

@article{huang2026identifying,
author={Huang, Jinyang and Feng, Yuanhao and Cui, Feng-Qi and Zhang, Xiang and Liu, Zhi and Liu, Xin and Liu, Jianchun and Zhang, Fusang and Li, Meng},
  journal={IEEE Transactions on Dependable and Secure Computing}, 
  title={Identifying Who You Are No Matter What You Write Through Abstracting Handwriting Style}, 
  year={2026},
  volume={23},
  number={3},
  pages={6890-6905},
  doi={10.1109/TDSC.2026.3668275}
}

@article{essahraui2026comprehensive,
  title={A comprehensive survey of tinyml-based biometric recognition for iot edge devices},
  author={Essahraui, Siham and Lamaakal, Ismail},
  journal={IEEE Internet of Things Journal},
  year={2026},
  publisher={IEEE}
}

@article{liang2020behavioral,
  title={Behavioral biometrics for continuous authentication in the internet-of-things era: An artificial intelligence perspective},
  author={Liang, Yunji and Samtani, Sagar and Guo, Bin and Yu, Zhiwen},
  journal={IEEE Internet of Things Journal},
  volume={7},
  number={9},
  pages={9128--9143},
  year={2020},
  publisher={IEEE}
}

\begin{IEEEbiography}
[{\includegraphics[width=1in,height=1.25in,clip,keepaspectratio]{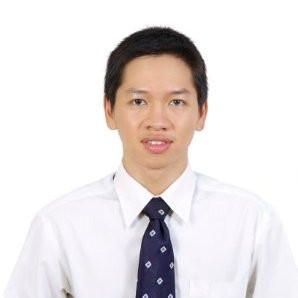}}]{Long Dang}
received the M.S. degree in Computer Science from the University of South Florida (USF), Tampa, FL, USA. He is currently pursuing the Ph.D. degree in Computer Science and Engineering at USF. His research interests include machine learning security, adversarial machine learning, data privacy, and trustworthy artificial intelligence, with a focus on securing machine learning systems against adversarial and data poisoning attacks.
\end{IEEEbiography}

\begin{IEEEbiography}
[{\includegraphics[width=1in,height=1.25in,clip,keepaspectratio]{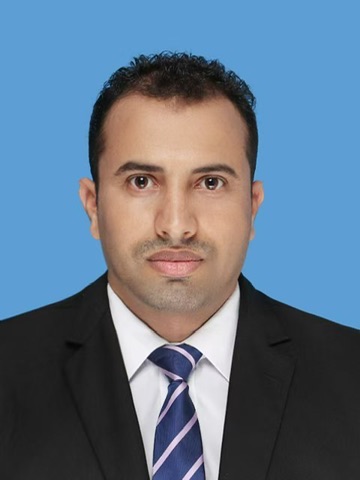}}]{Ali Alfatemi}
received the M.S. degree in computer science from South China University of Technology in 2021. He is currently a Ph.D. candidate and Teaching Fellow with the Department of Computer Science, Fordham University, New York, NY, USA. His research interests include AI for network security, robust and data-efficient machine learning, computer vision, large language models, and multimodal systems.
\end{IEEEbiography}

\begin{IEEEbiography}
[{\includegraphics[width=1in,height=1.25in,clip,keepaspectratio]{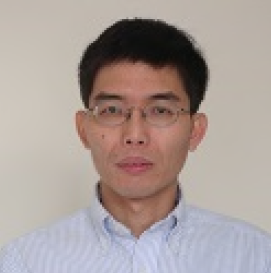}}]{Kaiqi Xiong (Senior Member, IEEE)}
received the Ph.D. degree in computer science from the North Carolina State University and the Ph.D. degree in mathematics from the Claremont Graduate School. Before returning to academia, he was with IT industry for several years. He is currently a Professor in the Intelligent Computer Networking and Security Laboratory with the University of South Florida (USF), affiliated with the Florida Center for Cybersecurity, the Department of Mathematics and Statistics, and the Department of Electrical Engineering. His research was supported by the National Science Foundation (NSF), NSF/BBN, the Air Force Research Laboratory, Amazon AWS, the Florida Center for Cybersecurity, and the Office of Naval Research. His research interests include machine learning/deep learning, security, and networking with their applications to various real-world computing systems. He received the Best Demo Award at the 22nd GENI Engineering Conference and the U.S. Ignite Application Summit with his team in 2015. He also received the best paper award at several conferences and the Global Research Award from the USF. He is a Senior Member of IEEE and ACM.
\end{IEEEbiography}

\begin{IEEEbiography}
[{\includegraphics[width=1in,height=1.25in,clip,keepaspectratio]{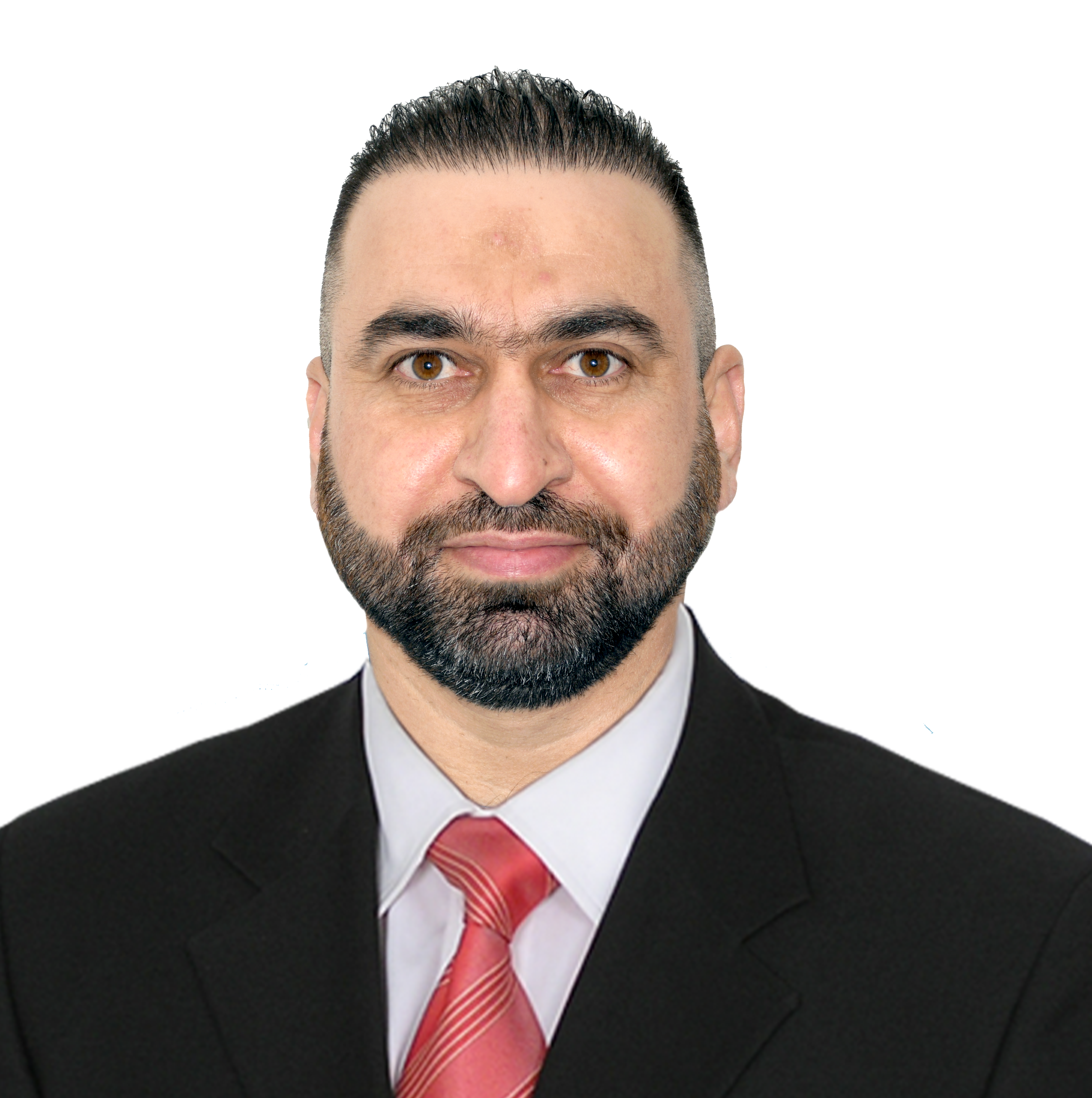}}]{Mohammed Aledhari (Senior Member, IEEE)} 
received his Ph.D. in computer science from Western Michigan University, Kalamazoo, MI, USA, in 2018. He is currently an Associate Professor in the Department of Data Science and the Department of Computer Science and Engineering at the University of North Texas, Denton, TX, USA, where he founded and directs the Computational Healthcare and Biotechnology Laboratory. His research interests include artificial intelligence, physical AI, machine learning, computational medicine, and wearable biosensors. Dr. Aledhari has been named to the Stanford/Elsevier list of the World's Top 2\% Most-Cited Scientists.
\end{IEEEbiography}

\begin{IEEEbiography}[{\includegraphics[width=1in,height=1.25in,clip,keepaspectratio]{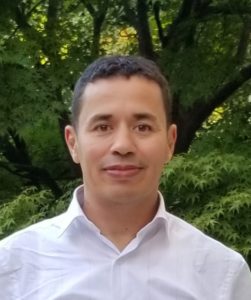}}]{Abdellah Chehri} (Senior Member, IEEE) is a Full Professor in the Department of Mathematics and Computer Science at the Royal Military College of Canada (RMC) in Kingston, Ontario. He also holds affiliate professorships at the Université du Québec en Outaouais (UQO) and the Université du Québec à Chicoutimi (UQAC), as well as an adjunct professorship at the University of Ottawa. His research spans a wide spectrum of advanced technologies, including intelligent mobility systems, sustainable transportation infrastructure, and vehicular communications. Dr. Chehri is internationally recognized for his contributions to V2X communications, 5G/6G and next-generation wireless networks, and AI-driven edge computing. His work integrates the Internet of Things (IoT), RF sensing, and signal processing to enable real-time decision-making in autonomous systems and smart environments. Dr. Chehri has served as a guest and associate editor for numerous highly regarded academic journals. He is actively involved in several professional societies, including the IEEE Public Safety Technology (PST) Initiative—where he serves as co-chair of the Transportation Committee—as well as the IEEE Communications Society (ComSoc), IEEE Vehicular Technology Society (VTS), and IEEE Photonics Society.
\end{IEEEbiography}

\begin{IEEEbiography}
[{\includegraphics[width=1in,height=1.25in,clip,keepaspectratio]{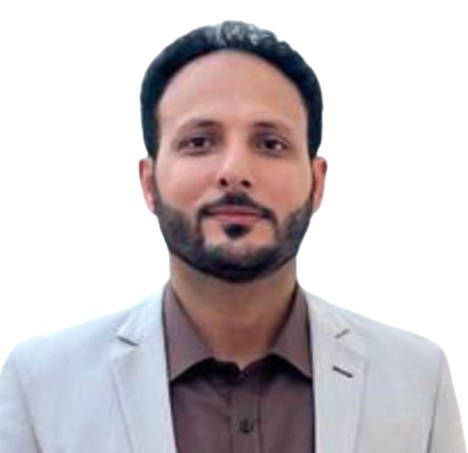}}]{Shawqi Al-Maliki}
received the M.Sc. degree in computer science from King Fahd University of Petroleum and Minerals (KFUPM), Dhahran, Saudi Arabia, in 2018, and the Ph.D. degree in computer science from Hamad Bin Khalifa University (HBKU), Doha, Qatar, in 2023. He is currently a Postdoctoral Researcher with HBKU. He was a Full-Stack Web Developer and Project Manager at Alyaum Media House, Dammam, Saudi Arabia, from 2009 to 2019, and a Computer Instructor at New Horizon Institute, Industrial Jubail, Saudi Arabia, from 2006 to 2008. His research interests include adversarial robustness, LLM security, privacy, reliability, and human-LLM alignment.
\end{IEEEbiography}

\begin{IEEEbiography}
[{\includegraphics[width=1in,height=1.25in,clip,keepaspectratio]{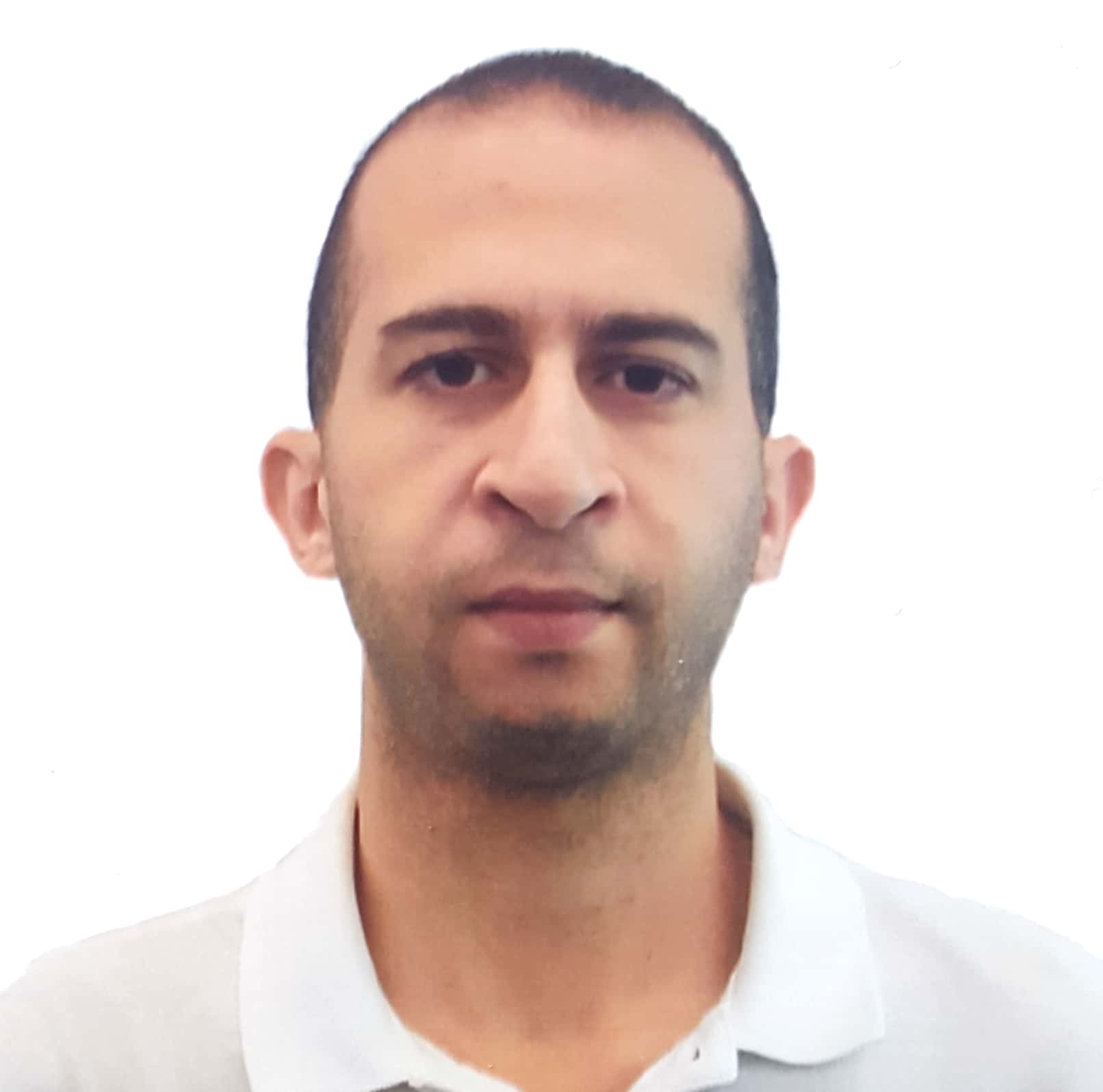}}]{Mohamed Rahouti}
received the Ph.D. degree in electrical engineering from the University of South Florida, Tampa, FL, USA, in 2020. He is currently an Associate Professor with the Department of Computer and Information Science, Fordham University, New York, NY, USA. His research has been supported by the U.S. National Science Foundation, the Florida Center for Cybersecurity, and the Qatar Research, Development, and Innovation Council. His research interests include computer networks and security, blockchain technologies, and artificial intelligence. Dr. Rahouti has served on the technical program committees of numerous conferences and workshops and currently serves in associate editor and guest editor roles for several peer-reviewed journals.
\end{IEEEbiography}

\end{document}